\documentclass[twocolumn]{aastex631}

\newcommand\J{CHIME J1634+44}
\newcommand\ILT{ILT J1101+5521}
\newcommand\GLEAM{GLEAM-X J0704-37}
\newcommand\ARSCO{AR Scorpii}
\newcommand\pelisoliWD{J191213.72--441045.1}
\newcommand{\red}[1]{\textcolor{black}{#1}}
\usepackage{mathtools}
\usepackage{csvsimple}

\begin{document}

\title{CHIME/FRB Discovery of an Unusual \red{Circularly Polarized} Long-Period Radio Transient with an Accelerating Spin Period}
\author[0000-0003-4098-5222]{Fengqiu Adam Dong}
  \affiliation{National Radio Astronomy Observatory, 520 Edgemont Rd, Charlottesville, VA 22903, USA}
  \affiliation{Green Bank Observatory, 155 Observatory Road, WV 24944, USA }
\author[0000-0002-6823-2073]{Kaitlyn Shin}
  \affiliation{MIT Kavli Institute for Astrophysics and Space Research, Massachusetts Institute of Technology, 77 Massachusetts Ave, Cambridge, MA 02139, USA}
  \affiliation{Department of Physics, Massachusetts Institute of Technology, 77 Massachusetts Ave, Cambridge, MA 02139, USA}
\author[0000-0002-4119-9963]{Casey Law}
  \affiliation{Cahill Center for Astronomy and Astrophysics, MC 249-17 California Institute of Technology, Pasadena CA 91125, USA}
  \affiliation{Owens Valley Radio Observatory, California Institute of Technology, Big Pine CA 93513, USA}
\author[0000-0002-0940-6563]{Mason Ng}
  \affiliation{Department of Physics, McGill University, 3600 rue University, Montr\'eal, QC H3A 2T8, Canada}
  \affiliation{Trottier Space Institute, McGill University, 3550 rue University, Montr\'eal, QC H3A 2A7, Canada}
  \affiliation{FRQNT Postdoctoral Fellow}
\author[0000-0001-9784-8670]{Ingrid Stairs}
  \affiliation{Department of Physics and Astronomy, University of British Columbia, 6224 Agricultural Road, Vancouver, BC V6T 1Z1 Canada}
\author[0000-0003-4056-9982]{Geoffrey Bower}
  \affiliation{Academia Sinica Institute of Astronomy and Astrophysics, 645 N. A'ohoku Pl., Hilo, HI 96720, USA}
  \affiliation{East Asian Observatory, 660 N. A'ohoku Pl., Hilo, HI 96720, USA}
\author[0009-0007-0757-9800]{Alyssa Cassity}
  \affiliation{Department of Physics and Astronomy, University of British Columbia, 6224 Agricultural Road, Vancouver, BC V6T 1Z1 Canada}
\author[0000-0001-8384-5049]{Emmanuel Fonseca}
  \affiliation{Department of Physics and Astronomy, West Virginia University, PO Box 6315, Morgantown, WV 26506, USA }
  \affiliation{Center for Gravitational Waves and Cosmology, West Virginia University, Chestnut Ridge Research Building, Morgantown, WV 26505, USA}
\author[0000-0002-3382-9558]{B.~M.~Gaensler}
  \affiliation{Department of Astronomy and Astrophysics, University of California, Santa Cruz, 1156 High Street, Santa Cruz, CA 95060, USA}
  \affiliation{Dunlap Institute for Astronomy and Astrophysics, 50 St. George Street, University of Toronto, ON M5S 3H4, Canada}
  \affiliation{David A. Dunlap Department of Astronomy and Astrophysics, 50 St. George Street, University of Toronto, ON M5S 3H4, Canada}
\author[0000-0003-2317-1446]{Jason W.~T.~Hessels}
  \affiliation{Department of Physics, McGill University, 3600 rue University, Montr\'eal, QC H3A 2T8, Canada}
  \affiliation{Trottier Space Institute, McGill University, 3550 rue University, Montr\'eal, QC H3A 2A7, Canada}
  \affiliation{Anton Pannekoek Institute for Astronomy, University of Amsterdam, Science Park 904, 1098 XH Amsterdam, The Netherlands}
  \affiliation{ASTRON, Netherlands Institute for Radio Astronomy, Oude Hoogeveensedijk 4, 7991 PD Dwingeloo, The Netherlands}
\author[0000-0001-9345-0307]{Victoria M.~Kaspi}
  \affiliation{Department of Physics, McGill University, 3600 rue University, Montr\'eal, QC H3A 2T8, Canada}
  \affiliation{Trottier Space Institute, McGill University, 3550 rue University, Montr\'eal, QC H3A 2A7, Canada}
\author[0009-0008-6166-1095]{Bikash Kharel}
  \affiliation{Department of Physics and Astronomy, West Virginia University, PO Box 6315, Morgantown, WV 26506, USA }
  \affiliation{Center for Gravitational Waves and Cosmology, West Virginia University, Chestnut Ridge Research Building, Morgantown, WV 26505, USA}
\author[0000-0002-4209-7408]{Calvin Leung}
  \affiliation{Miller Institute for Basic Research, University of California, Berkeley, CA 94720, United States}
  \affiliation{Department of Astronomy, University of California, Berkeley, CA 94720, United States}
\author[0000-0002-7164-9507]{Robert A.~Main}
  \affiliation{Department of Physics, McGill University, 3600 rue University, Montr\'eal, QC H3A 2T8, Canada}
  \affiliation{Trottier Space Institute, McGill University, 3550 rue University, Montr\'eal, QC H3A 2A7, Canada}
\author[0000-0002-4279-6946]{Kiyoshi W.~Masui}
  \affiliation{MIT Kavli Institute for Astrophysics and Space Research, Massachusetts Institute of Technology, 77 Massachusetts Ave, Cambridge, MA 02139, USA}
  \affiliation{Department of Physics, Massachusetts Institute of Technology, 77 Massachusetts Ave, Cambridge, MA 02139, USA}
\author[0000-0002-2885-8485]{James W.~McKee}
  \affiliation{Department of Physics and Astronomy, Union College, Schenectady, NY 12308, USA}
\author[0000-0001-8845-1225]{Bradley W.~Meyers}
  \affiliation{International Centre for Radio Astronomy Research (ICRAR), Curtin University, Bentley WA 6102, Australia}
  \affiliation{Australian SKA Regional Centre (AusSRC), Curtin University, Bentley WA 6102, Australia}
\author[0009-0005-0466-9371]{Obinna Modilim}
  \affiliation{MIT Kavli Institute for Astrophysics and Space Research, Massachusetts Institute of Technology, 77 Massachusetts Ave, Cambridge, MA 02139, USA}
\author[0000-0002-8897-1973]{Ayush Pandhi}
  \affiliation{David A. Dunlap Department of Astronomy and Astrophysics, 50 St. George Street, University of Toronto, ON M5S 3H4, Canada}
  \affiliation{Dunlap Institute for Astronomy and Astrophysics, 50 St. George Street, University of Toronto, ON M5S 3H4, Canada}
\author[0000-0002-8912-0732]{Aaron B.~Pearlman}
  \affiliation{Department of Physics, McGill University, 3600 rue University, Montr\'eal, QC H3A 2T8, Canada}
  \affiliation{Trottier Space Institute, McGill University, 3550 rue University, Montr\'eal, QC H3A 2A7, Canada}
  \affiliation{Banting Fellow}
  \affiliation{McGill Space Institute Fellow}
  \affiliation{FRQNT Postdoctoral Fellow}
\author[0000-0001-5799-9714]{Scott M.~Ransom}
  \affiliation{National Radio Astronomy Observatory, 520 Edgemont Rd, Charlottesville, VA 22903, USA}
\author[0000-0002-7374-7119]{Paul Scholz}
  \affiliation{Department of Physics and Astronomy, York University, 4700 Keele Street, Toronto, ON MJ3 1P3, Canada}
  \affiliation{Dunlap Institute for Astronomy and Astrophysics, 50 St. George Street, University of Toronto, ON M5S 3H4, Canada}
\author[0000-0002-2088-3125]{Kendrick Smith}
  \affiliation{Perimeter Institute of Theoretical Physics, 31 Caroline Street North, Waterloo, ON N2L 2Y5, Canada}
\newcommand{\allacks}{

F.A.D. is supported by a Jansky Fellowship.
K.S. is supported by the NSF Graduate Research Fellowship Program.
C.J.L. has been supported by NSF award 2022546.
M.N. is a Fonds de Recherche du Québec – Nature et Technologies (FRQNT) postdoctoral fellow.
Pulsar and FRB research at UBC is supported by an NSERC Discovery Grant and by the Canadian Institute for Advanced Research. 
E.F. is supported by the National Science Foundation under grant AST-2407399.
J.W.T.H. and the AstroFlash research group acknowledge support from a Canada Excellence Research Chair in Transient Astrophysics (CERC-2022-00009).
V.M.K. holds the Lorne Trottier Chair in Astrophysics \& Cosmology, a Distinguished James McGill Professorship, and receives support from an NSERC Discovery grant (RGPIN 228738-13).
C. L. is supported by the Miller Institute for Basic Research at UC Berkeley.
K.W.M. holds the Adam J. Burgasser Chair in Astrophysics.
A.P. is funded by the NSERC Canada Graduate Scholarships -- Doctoral program.
A.B.P. is a Banting Fellow, a McGill Space Institute~(MSI) Fellow, and a Fonds de Recherche du Quebec -- Nature et Technologies~(FRQNT) postdoctoral fellow.
The National Radio Astronomy Observatory is a facility of the National Science Foundation operated under cooperative agreement by Associated Universities, Inc. SMR is a CIFAR Fellow and is supported by the NSF Physics Frontiers Center award 2020265.
P.S. acknowledges the support of an NSERC Discovery Grant (RGPIN-2024-06266).

We acknowledge that CHIME is located on the traditional, ancestral, and unceded territory of the Syilx/Okanagan people. We are grateful to the staff of the Dominion Radio Astrophysical Observatory, which is operated by the National Research Council of Canada.  CHIME is funded by a grant from the Canada Foundation for Innovation (CFI) 2012 Leading Edge Fund (Project 31170) and by contributions from the provinces of British Columbia, Qu\'ebec and Ontario. The CHIME/FRB Project, which enabled development in common with the CHIME/Pulsar instrument, is funded by a grant from the CFI 2015 Innovation Fund (Project 33213) and by contributions from the provinces of British Columbia and Qu\'{e}bec, and by the Dunlap Institute for Astronomy and Astrophysics at the University of Toronto. Additional support was provided by the Canadian Institute for Advanced Research (CIFAR), McGill University and the McGill Space Institute thanks to the Trottier Family Foundation, and the University of British Columbia. The CHIME/Pulsar instrument hardware was funded by NSERC RTI-1 grant EQPEQ 458893-2014.

This research was enabled in part by support provided by the BC Digital Research Infrastructure Group and the Digital Research Alliance of Canada (alliancecan.ca).

The National Radio Astronomy Observatory and Green Bank Observatory are facilities of the U.S. National Science Foundation operated under cooperative agreement by Associated Universities, Inc.

This work made use of data supplied by the UK Swift Science Data
Centre at the University of Leicester and the Swift satellite. Swift,
launched in November 2004, is a NASA mission in partnership with
the Italian Space Agency and the UK Space Agency. Swift is managed
by NASA Goddard. Penn State University controls science and flight
operations from the Mission Operations Center in University Park,
Pennsylvania. Los Alamos National Laboratory provides gamma-ray
imaging analysis.
}

\begin{abstract}
\red{We report the discovery of CHIME J1634+44, a Long Period Radio Transient (LPT) unique for two aspects: it is the first known LPT to emit fully circularly polarized radio bursts, and it is the first LPT with a significant spin-up.} Given that high circular polarization ($>90$\%) has been observed in FRB~20201124A and in some giant pulses of PSR~B1937+21, we discuss the implications of the high circular polarization of CHIME J1634+44 and conclude its emission mechanism is likely to be ``pulsar-like''. While CHIME J1634+44 has a pulse period of 841\,s, its burst arrival patterns are indicative of a secondary 4206\,s period, probably associated with binary activity. The timing properties suggest it has a significantly negative period derivative of $\dot{P}=-9.03(0.11) \times 10^{-12}$ \,s\,s$^{-1}$. \red{Few systems have been known to spin-up, most notably transitional millisecond pulsars and cataclysmic binaries, both of which seem unlikely progenitors for CHIME J1634+44.} If the period was only associated with the spin of the object, then the spin up is likely generated by accretion of material from a companion. If, however, the radio pulse period and the orbital period are locked, \red{as appears to be the case for two other LPTs}, the spin up of CHIME J1634+44 could be driven by gravitational wave radiation.
\end{abstract}



\section{Introduction} \label{sec:intro}
In recent years, a new class of ``Long Period Radio Transients'' (LPTs) have been discovered, with periods ranging from tens of seconds to hours \citep{Hurley-Walker:Zhang:Bahramian:2022,Caleb:2022,Hurley-Walker:2023,Caleb:2024,Dong:2024,deRuiter:2024,Hurley-Walker:2024,Wang:Rea:Bao:2024,Li:Yuan:Wu:2024,lee:caleb:2025}. LPTs have been named as such due to their long radio periods compared to pulsars, while still producing coherent emission. The emission of LPTs is required to be coherent as these sources are bright and detected at low frequencies (less than 300~MHz in some cases; \citealt{Hurley-Walker:Zhang:Bahramian:2022,Hurley-Walker:2023}), implying a non-physical brightness temperature \citep{Manchester:1977,Hurley-Walker:Zhang:Bahramian:2022}.

Two models have arisen as likely progenitors for LPTs --- magnetic white dwarfs (WDs), and neutron stars (NSs) \citep{Katz:2022,Beniamini:2023,Rea:2024}. These models are favored due to two classes of known astrophysical objects that produce coherent radio emission, WD pulsars and NS pulsars.  In the LPT population, we find evidence for both source types. For example, two LPTs, ILT J1101+5521 and GLEAM-X J0704--37, are WDs in a binary orbit with M dwarf stars, confirmed via optical spectroscopy \citep{deRuiter:2024,Hurley-Walker:2024,Rodriguez:2025}. \red{WD binary LPTs pulse on periods of hours, much longer than the rotation period of WD pulsars\citep[minutes;][]{Marsh:Gansicke:hummerich:2016,Buckley:2017,Pelisoli:2023}. In fact, WD binary LPTs pulse on periods similar to that of the orbit of WD binary pulsar systems}. Other LPTs, like PSR J0901-4046, are most likely isolated NSs. This is demonstrated through a stable pulse profile, indicating persistent beamed emission like a NS pulsar \citep{Caleb:2022}. 
It is becoming increasingly clear that there may be two different populations of LPTs, namely those that exist in WD binaries and those that are slowly rotating neutron stars.

Here, we report the discovery of a new LPT discovered by the Canadian Hydrogen Intensity Mapping Experiment (CHIME)/Fast Radio Burst (FRB)/Pulsar survey for Galactic pulsars, \J{}. This source was \red{independently} co-discovered during the LOFAR Two-meter Sky Survey Data Release 2 \citep{shimwell:2019,shimwell:2022} as part of a Stokes V transient search (Bloot et al., 2025). In Section \ref{sec:discovery}, we detail the telescopes and modes of operation that were used. Section \ref{sec:vlapol} details the measurements made by the VLA. Section \ref{sec:optical} details a search for optical counterparts. Section \ref{sec:timing} provides a long baseline timing solution and poses several conundra in \J{}'s timing properties. Finally, we discuss the results and possible implications of the \J{} system in Section \ref{sec:discussion}.

\section{Discovery and Observations} \label{sec:discovery}
\begin{figure}[ht!]
\includegraphics[width=\linewidth]{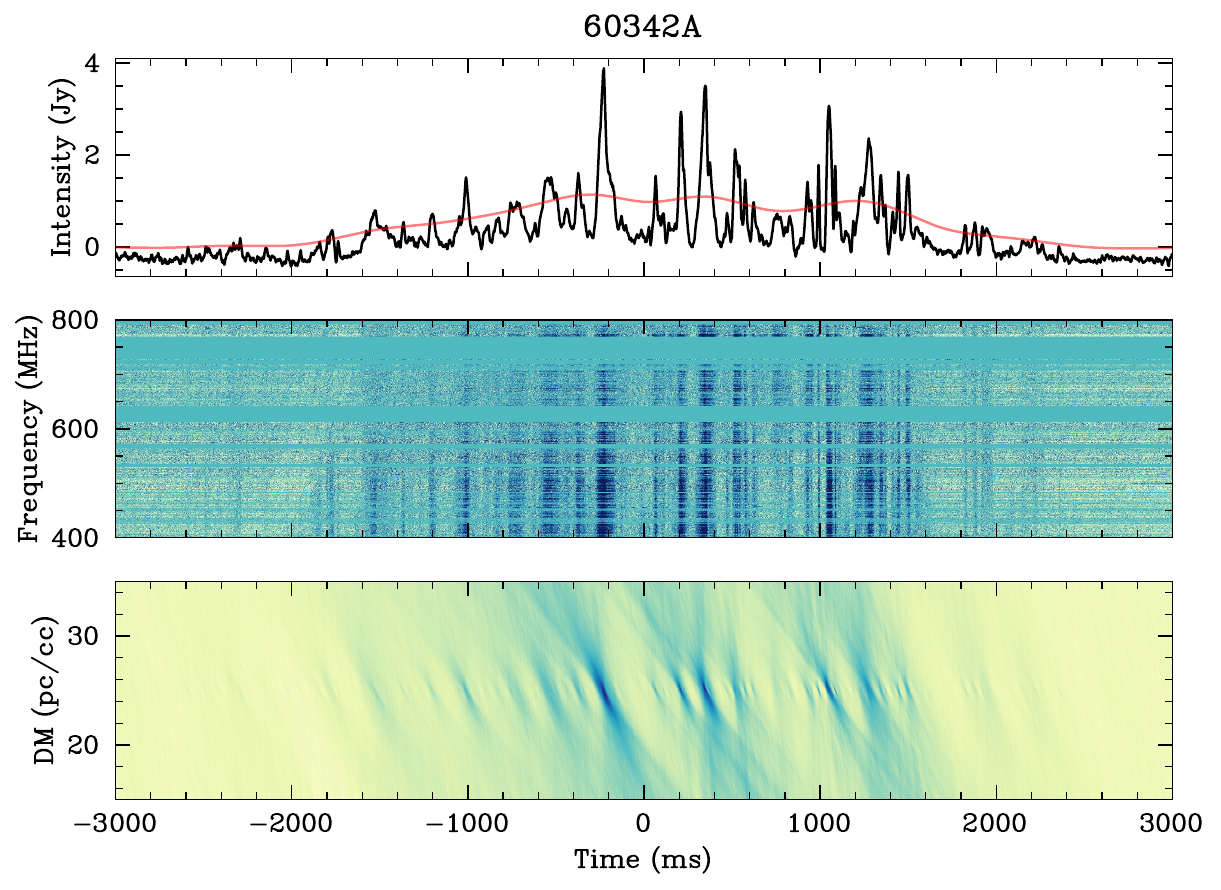}
\caption{Sample of a detection made with the CHIME/Pulsar instrument. The top panel shows the \red{dedispersed} flux-calibrated frequency-averaged profile. The red line is the smoothed profile for TOA extraction, the middle panel shows the \red{dedispersed dynamic spectrum}, and the bottom panel shows the DM-time power spectrum. The title is the MJD the burst was detected, following the convention in \cite{Dong:2024}.}
\label{fig:example_det}
\end{figure}
CHIME is a transit radio telescope in British Columbia, Canada. It consists of four 20x100\,m cylindrical dishes. The unique cylindrical design allows for a wide instantaneous field of view of $\sim$200 square degrees over the 400-800\,MHz bandwidth in observing frequency. While CHIME's primary goal is to detect hydrogen at cosmological distances \citep{10.3847/1538-4365/ac6fd9}, the large spatial and frequency coverage enables CHIME to perform rapid all-sky daily surveys for pulsars, FRBs, and other transient events. CHIME, therefore, conducts multiple commensal experiments like CHIME/FRB \citep{10.3847/1538-4357/aad188} and CHIME/Pulsar \citep{10.3847/1538-4365/abfdcb}.

\J{} was discovered in the CHIME/FRB single pulse pulsar survey, where we are using the CHIME/FRB trigger criteria \citep{10.3847/1538-4357/aad188} for all sources with a dispersion measure (DM) low enough to be considered inside the Milky Way Galaxy according to both the NE2001 and the YMW16 DM models \citep{Cordes:2001,Yao:Manchester:2016}. This survey utilizes the full observing capabilities of the 1024 Fast Fourier Transformed beams to survey the CHIME sky \citep{10.3847/1538-4357/aad188}. We discovered the first burst from \J{} on MJD 59883 (2022 October 31). We observed multiple instances of reactivation after that, beginning on MJD 60003 (2023 February 28) and then again for a longer activation timespan on MJD 60270 (2023 November 22). During this timespan, we identified a period of $\sim$841\,s. A search through archival metadata of CHIME/FRB revealed that there were bursts from \J{} dating back to MJD\,58893 (2022 February 14). These bursts only have trigger information, such as timestamps, signal-to-noise (S/N), and DM. We can confirm that they are indeed astrophysical as they phase connect with \J{} bursts that have been verified. Activity epochs of \J{} were sporadic, but we retrieved channelized raw voltage data (hereafter referred to as baseband data, \citealt{10.3847/1538-4357/ad464b}) after MJD 60270 (2023 November 22).

CHIME/Pulsar tracking beam observations record 327.68\,$\mu$s dynamic spectra in SIGPROC filterbank format \footnote{\url{https://sigproc.sourceforge.net/}}. The first detection made with CHIME/Pulsar was on MJD 60270 (2023 December 23). An example is shown in Figure \ref{fig:example_det}\footnote{The rest of the detection plots, including those by CHIME/FRB, are provided at \url{https://github.com/CHIMEFRB/J1634-44_additional_information}}.


The sudden active state of \J{} on MJD 60270 (2023 November 22) resulted in Director's Discretionary Time (DDT) observations with the Karl G. Jansky Very Large Array (VLA), utilizing the \textit{realfast} system at 10~ms integrations and visibility data at 2~s integrations for 4 hours (23B-337). We also triggered Target of Opportunity (ToO) observations using the {\it Neil Gehrels Swift X-ray Telescope (Swift)} for 10~ks (target ID 16412) and the Green Bank Telescope (GBT) for 4 hours at 680-920~MHz. Both the GBT and the VLA detected two bursts. The VLA/\textit{realfast} detections led to a localization of (RA, Dec) = (16h34m29.96s$\pm0.5\arcsec$, +44d50m13.5s$\pm1.1\arcsec$)\footnote{All coordinates are given in the J2000 reference frame.}. \red{The {\it Swift XRT} data showed no sources within the 3-$\sigma$ localization error region; however, we place an upper limit on the 0.3--10\,keV luminosity $L_{X}<1.3-5.2\times10^{32}{\rm\,ergs\,s^{-1}}$ depending on the model used. Details are provided in Appendix \ref{sec:xray}.}

To summarize all of the detections, the most extensive set was made with the CHIME/FRB and CHIME/Pulsar instruments, which detected 69 and 44 bursts, respectively. The two instruments frequently detected the same bursts. The GBT and the VLA detected 2 bursts each. In total, 89 unique bursts were detected over a $\sim4.5$\,yr period.

We calculate the quasiperiod on all bursts that show significant microstructure via an autocorrelation function and a Fast Fourier Transform (FFT). the details are presented in Appendix \ref{sec:quasiperiodicity}. Quasiperiodicity indeed exists among many bursts emitted by \J{} with a median quasiperiod of $P_{\mu}=0.13(11)$\,s. We find that this does not align well with the universal NS quasiperiod relation described by \cite{kramer:liu:2023}.

We flux calibrated all bursts from CHIME/Pulsar using three calibrators which bracket the \J{} declination range. They are 3C380, 3C196, and NGC1265 at declinations of +48.75$^\circ$, +48.22$^\circ$, and +41.9$^\circ$, respectively. The procedure is described in Appendix \ref{sec:flux} and \cite{Dong_2024}. The dispersion measure is calculated via variance optimization of the DM-time power spectrum and further describe in Appendix \ref{sec:discovery_dm}.

\section{Very Large Array}
\label{sec:vlapol}
The \textit{realfast} search system triggered on two unique bursts from \J{} at MJD 60301.76197 and 60301.85935 topocentric time.
The detection was made at the edge of the triggered buffer window for both bursts, so we can only provide a lower limit on the temporal width of the bursts of roughly 3 seconds. Due to the temporal width and higher observing frequency, the DM is poorly constrained compared to the CHIME measurements.

\begin{figure*}
\centering
\includegraphics[width=\textwidth]{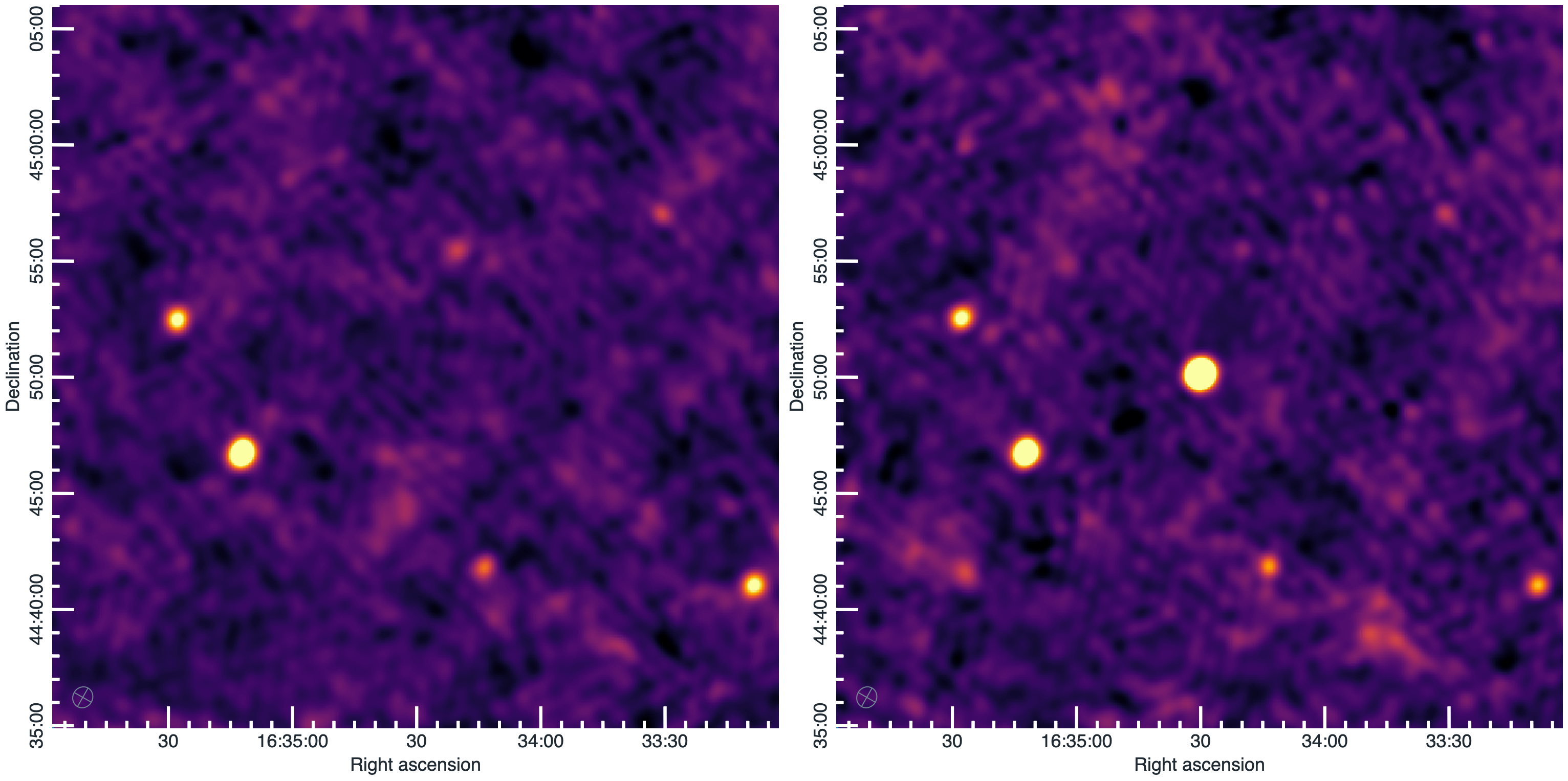}
\caption{VLA radio continuum images of the LL and RR correlation products toward \J{}. The images were made from visibilities selected from the entire \red{9\,s triggered data} time span and frequencies from 1328 to 1456 MHz. The source is only significantly detected in RR polarization, which is consistent with 100\% circular polarization.}
\label{fig:vlarrll}
\end{figure*}

We calibrated and imaged the 9\,s fast visibility data in the band with the strongest burst emission (128 MHz centered at 1392 MHz) to measure the burst polarization and the source position. The triggered recording only includes the two auto-correlation products, RR and LL. As shown in Figure \ref{fig:vlarrll}, the RR and LL polarization images show that \J{} differs from all other sources in the image in that it is only detected in the RR correlation product. We find the source to be spatially unresolved with a mean flux of 77$\pm$1\ mJy in the RR correlation product. In the LL product, we measure an upper limit of 1.8 mJy (3 sigma). \red{This implies a lower limit on the circular polarization fraction of 98\%. We provide the specific technical details of the VLA observations in Appendix \ref{subsec:discovery_VLA}}

\subsection{Astrometry}
\label{subsec:vla_astrometry}

We find a best-fit source position with a statistical error of 0.3\arcsec{} using the RR image. However, the absolute astrometry of the VLA is typically limited by gain calibration to roughly 1/20th of the synthesized beam size of 46\arcsec{}\citep{1997ApJ...475..479W}. To improve on this, we directly measure the astrometric accuracy for this image by matching radio sources in the field to sources cataloged in the FIRST survey \cite[version 2014Dec17;][]{2015ApJ...801...26H}. Since FIRST has an astrometric error of 20 milliarcsec, we expect the astrometry of \J{} to be limited by the statistical errors in the primary measurement.


Four of the seven nearest radio sources seen in Figure \ref{fig:vlarrll} have unresolved counterparts in FIRST. To make the \J{} astrometry consistent with FIRST, we must apply a mean shift of ($\Delta$RA, $\Delta$Dec) $= (-0.7, 2.7)\arcsec{}$. After applying this shift, we find the position of \J{} to be (RA, Dec) = (16h34m29.96s +44d50m13.5). Treating the scatter of the cross-matched sources as an estimate of the astrometric precision, we find a total position $1\sigma$ uncertainty of $(0.5\arcsec, 1.1\arcsec)$. These errors should be treated as an upper limit on the astrometric precision caused by cross-matched calibration sources; the lower limit is the statistical error in the primary measurement of 0.3\arcsec. 
\section{Optical}\label{sec:optical}

\begin{figure}[ht]
    \centering
    \includegraphics[width=\linewidth]{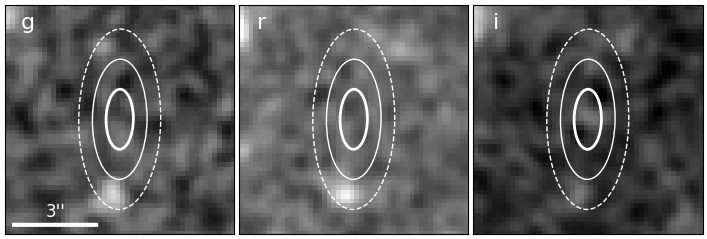}
    \caption{Archival optical imaging from HSC, smoothed and centered on the VLA/\textit{realfast} localization for \J{}; up is North and left is East.
    The \textit{g}-band image from the PDR3, and the \textit{r}- and \textit{i}-band images are from the HSCLA.
    Overlaid on the images are 1-$\sigma$, 2-$\sigma$, and 3-$\sigma$ contours in bold, solid, and dashed lines.
    The possible optical source is at the bottom of the 3-$\sigma$ contours, most visible in the \textit{g}- and \textit{r}-band images.
    }
    \label{fig:hsc_gri}
\end{figure}

Due to the presence of optical counterparts to some discovered LPTs \citep{deRuiter:2024,Hurley-Walker:2024}, we searched through archival data to determine the presence of possible optical counterparts to \J{}.
Public data on this field of view are available from the Hyper Suprime-Cam (HSC) via the HSC Subaru Strategic Program third public data release \citep[HSC-SSP PDR3;][]{2022PASJ...74..247A} and the HSC Legacy Archive \citep[HSCLA;][]{2021PASJ...73..735T}.
These are the deepest publicly available observations of this field.
Within the 3-$\sigma$, but outside the 2-$\sigma$ localization region determined by VLA/\textit{realfast} (Section~\ref{subsec:vla_astrometry}), one optical source is visible (Figure~\ref{fig:hsc_gri}) in HSC data. 

While HSC observes in \textit{grizy} filters, in the public data release, photometry was only successfully determined for this source in the \textit{g} and \textit{r} filters.
Visual inspection of the PDR3 coadded images confirms that this source is not significantly detected in redder filters.
In the HSCLA observations of this field, there appears to be a marginal source in the \textit{i} filter; however, it is too faint to extract reliable photometry.
The Kron magnitudes of this source are $m_{\mathrm{AB},g} = 25.3 \pm 0.4$ and $m_{\mathrm{AB},r} = 25.5 \pm 0.8$.
The 5-$\sigma$ limiting magnitudes around this source for the other filters are $m_{\mathrm{AB},i} = 25.4$, $m_{\mathrm{AB},z} = 24.4$, and $m_{\mathrm{AB},y} = 23.4$.

\red{We estimate a chance alignment probability by obtaining all the HSC PDR3 sources within a 3$^\circ$ radius around the VLA/\textit{realfast} localization of \J{}, that also have $g$-band and $r$-band photometry at least as bright as this optical candidate source.
We draw $10^6$ random positions within this region and determine the number of times a drawn position lies within 3.3\arcsec (the 3-$\sigma$ uncertainty in Dec) of an optical source; we find this occurs $\approx$5.8\% of the time.}

Given its classification by the HSC pipeline as extended rather than point-like, and the separation from the VLA localization, it is possible that this optical source is a background object (e.g., a background galaxy) rather than a true optical counterpart.
Even if we assume that the classification as an extended source is incorrect (plausible because of its faint magnitude), the possible interpretation of this optical source is relatively limited given our two faint photometric data points.
The optical source appears to be tentatively brighter in the $g$- band than the $r$- band, which may be promising for a WD progenitor given the rise of the spectral energy distribution towards bluer wavelengths; however, we did not find $u$-band data for this source, and there is no detection in the UV in archival \textit{GALEX} \citep{2005ApJ...619L...1M} data.
If this optical source were a true counterpart to \J{}, consistent WD temperatures could range from $\sim$20,000~K to $\sim$40,000~K.
\red{The optical source being brighter in $g$- band than in $r$- band is inconsistent with the expected spectral energy distribution of an M dwarf.
However, at the DM-inferred distance of \J{}, the presence of a faint late-M dwarf cannot be ruled out.}


\section{Timing} \label{sec:timing}

This section provides the methodology for the timing analysis of \J{}.
All the provided TOAs are topocentric at the position of the respective observatories. We use PINT \citep{luo:ransom:demorest:2021} for the barycentering, frequency correction, and the timing model fit of the data.
The TOA errors on each burst are large enough such that the clock offsets between the CHIME/FRB instrument and the CHIME/Pulsar instruments are insignificant in comparison \citep{Dong:2024}. Nevertheless, we found all coincident detections between CHIME/Pulsar and CHIME/FRB and fit the frequency, the first frequency derivative, and a global timing offset between the two datasets. The resultant offset is 0.3(3)\,s. For the rest of the timing analysis, we proceed with this value fixed as the timing offset between these two data sets. We provide the full timing solution in Table \ref{tab:timing_tables}.

\J{} was initially identified to possess a period of $\sim4206$~s (70.1 minutes). With this period and CHIME's short $\sim10$~minute transit, we expect to make a detection of \J{} every second day. This 2-day spacing is a beat period between 4206\,s and CHIME's observation cadence. This spin period appeared robust until MJD 60348, when we detected two bursts on \red{consecutive} days. This detection confirmed the need for a shorter period of $\sim$841\,s. \red{However, this posed other concerns. If the period was indeed 841\,s, why did most of our bursts occur with a 2-day spacing? We find that the number of bursts we detect with a 2-day separation (i.e., in agreement with 4206\,s) has a probability of occurring by random chance of $ 3.5\times10^{-9}$ (shown in Appendix \ref{sec:period_ambiguity}). This is discussed further in Section \ref{sec:discussion}}.
\begin{table}[ht]
  \caption{}
  \label{tab:timing_tables}
  \centering
  \begin{tabular}{ll}
    \hline
    \hline
    R.A. (hh:mm:ss)$^{*}$                    & 16h34m29.96s \\      
    Dec (dd:mm:ss)$^{*}$                     & 44$^{\circ}$50'13.5"\\
    $P$(s)                                   & 841.245895(6)     \\ 
    $\dot{P}(\times10^{-12}~\text{ss}^{-1})$ & --9.03(0.11)    \\   
    PEPOCH(MJD)                              & 60366           \\   
    TIMEEPH                                  & FB90            \\   
    NTOA                                     & 127             \\  
    CLOCK                                    & TT(BIPM2023)    \\  
    RMS Residuals (phase)                    & 0.0055          \\  
    RMS Residuals (s)                        & 4.6             \\  
    \hline
    Derived Values                           &                   \\ 
    \hline
    Galactic Longitude (deg)                 & 70.17             \\  
    Galactic Latitude (deg)                  & +42.58            \\ 
    \hline
  \end{tabular}\\ $^{*}$ These are fixed at the VLA positions.
\end{table}

\begin{figure*}[ht]
\centering
\includegraphics[width=\linewidth]{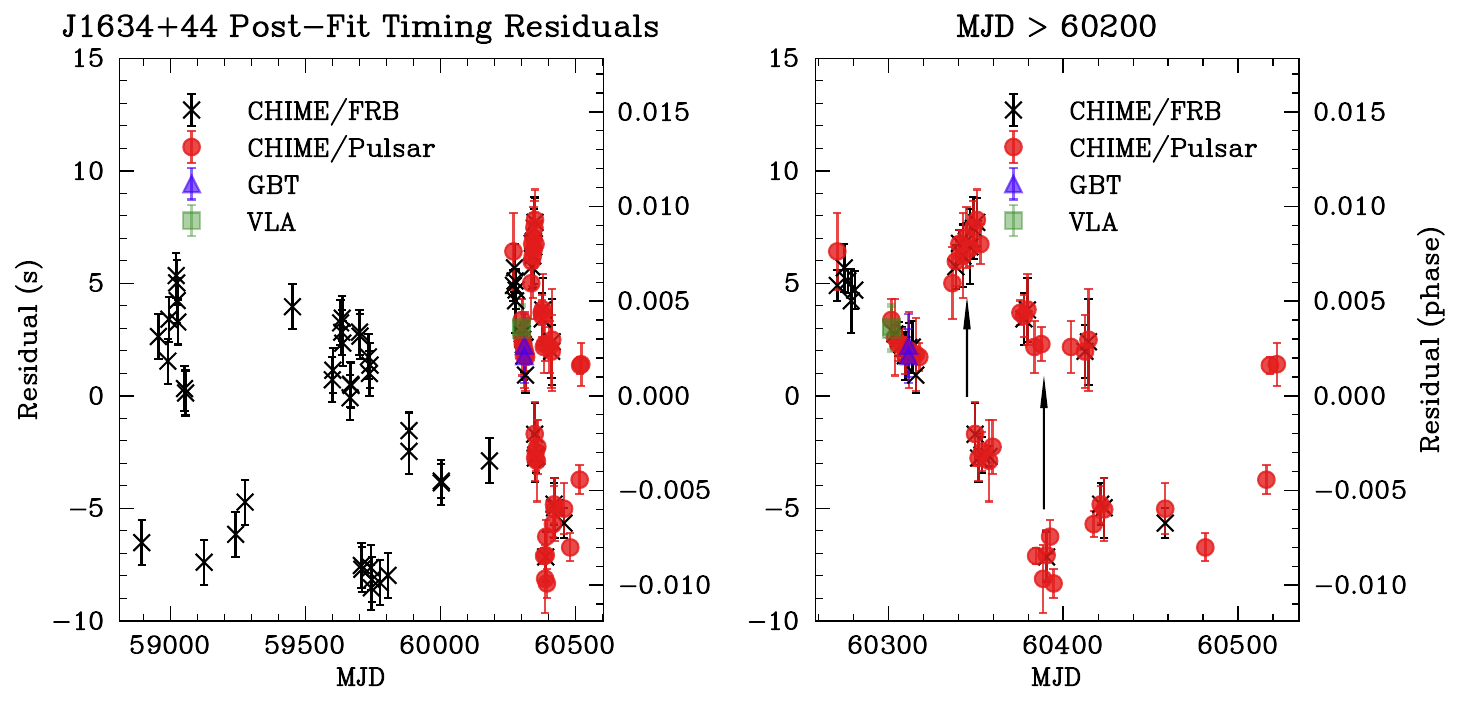}
\caption{Residuals for a 841~s period. \red{The error bars are 1$\sigma$}. The right panel shows a zoomed-in image of the densest portion of TOAs when all instruments were observing \J{}. The arrows show \red{offsets possibly due to timing noise between burst as discussed in the text.}}
\label{fig:841_timing}
\end{figure*}

With an 841~s period, the residuals show \red{some} scatter relative to the typical TOA error. \red{The residual scatter is about $2\%$ of the period, roughly similar to the duty cycles found in \cite{Hurley-Walker:2024} and \cite{deRuiter:2024}. However, the bursts from \J{} appear to be narrower than the aforementioned sources.} The overall timing residuals are still small compared to the period ($\lesssim|1|$\%). It also appears that there are some scatter \red{between bursting epochs in the data}. For example, see the right panel \red{arrows} in Figure \ref{fig:841_timing}. \red{This may be caused by timing noise}. The period derivative for \J{} is negative at $>80$-$\sigma$, indicating that there are \red{additional physics that the timing process does not capture, e.g., possible accretion of material onto \J{}}, as radiating objects are generally expected to lose rotational energy and spin down (increasing period). We discuss the possible implications of the negative period derivative in Section \ref{sec:discussion}.


\section{Discussion} \label{sec:discussion}

\subsection{Polarization}

The polarization of \J{} is rare among the radio emissions emitted by compact objects. It is challenging to produce 100\% circular polarization at the luminosities observed. Fully circularly polarized radio pulses or bursts have been observed in main sequence stars such as the Sun \citep{Dulk:1985}, CU~Vir \citep[e.g.,][]{lo:bray:2012}, M~dwarfs \citep[e.g, YZ Ceti, YZ~CMi;][]{pineda:villadsen:2023,rucinski:1994}, auroral (or magnetospheric) processes around cool brown dwarfs \citep{2007ApJ...663L..25H, 2008ApJ...684..644H}, magnetic cataclysmic variables \citep[e.g.][]{barrett:2020}, and the Jovian system \citep[e.g.,][]{Marques:Zarka:2017}. These are all close by (within 75\,pc) and produce pulses three orders of magnitude lower in flux than \J{}, resulting in five orders of magnitude lower luminosities. It is thus unlikely that \J{} shares the same underlying emission mechanisms as any of these objects.

There is strong evidence that some Fast Radio Bursts (FRBs) are produced by NSs through magnetar bursts \citep{10.1038/s41586-020-2863-y,Bochenek:Ravi:2020} or pulsar emission \citep{Mckinven:2024,nimmo:2025}. Recently, FRB 20201124A has been observed to produce $\sim$90\% circularly polarized bursts \citep{Jiang:Xu:Niu:2025}. Most pulsars have low circular polarization \citep{han:manchester:1998}. However, PSR B1937+21 has shown that the brightest giant pulses can be almost fully circularly polarized \citep{cognard:shrauner:1996}. Giant pulses are usually offset from the main radio components, as shown in the case of PSR B1937+21 \citep{cognard:shrauner:1996} and PSR B0540-69 \citep{Johnston:Romani:2004}. They are, however, aligned with the X-ray and/or $\gamma$--ray emission, suggesting a common origin \citep{Enoto:Terasawa:Kisaka:2021}. \red{Scaling PSR B1937+21 to the distance of \J{}, our X-ray detection threshold of $1.2\times10^{-13}$erg\,cm$^{-2}$\,s$^{-1}$ should have detected something PSR B1937+21-like \citep{cusumano:hermsen:kramer:2003}, but as discussed in Section \ref{sec:multiwavelength_cp}, X-ray pulsars can be much fainter than PSR B1937+21}. This is suggestive that \J{} may produce as-of-yet undetected high-energy pulsations. While the emission mechanisms for giant pulses remain highly debated, the mechanisms that generate "normal" coherent radiation from pulsars are likely curvature radiation and/or inverse Compton scattering of particle bunches \citep{Lorimer:Kramer:2012}. Both models require a line of sight off-center from the radiation axis to create the fully circularly polarized bursts we observe. Emission from particle bunches, particularly, can produce highly circularly polarized bursts despite being only slightly off-axis \citep{Jiang:Xu:Niu:2025}.

Regardless of the source, propagation effects could also cause circular polarization. For example, generalized Faraday rotation (also called Faraday conversion) can partially convert linearly polarized light to circularly polarized \citep{vedantham:ravi:2019,qu:zhang:2023}. However, this does not usually result in fully circularly polarized bursts as we have observed in \J{}. Additionally, these effects are frequency-dependent; the LOFAR-CHIME-VLA measurements (Bloot et al., 2025) show that fully circularly polarized bursts remain robust over a decade in observing frequencies. Therefore, we conclude that the circular polarization is intrinsic to \J{}.

\red{We conclude that ``pulsar-like'' emission mechanisms are the most likely for \J{} as fully circularly polarized bursts have been observed from NSs at the energetics required.}

\subsection{X-ray counterparts} \label{sec:multiwavelength_cp}
Multiwavelength detections of \J{} can enable further characterization of its origin. For example, the X-ray detection of DART/ASKAP~J1832$-$0911 may suggest it is an older ($\gtrsim0.5{\rm\,Myr}$) magnetar or a highly-magnetized WD in a WD-M dwarf binary (see \cite{Wang:Rea:Bao:2024}). Although the age may be debated due to the supernovae remnant coincident found by \cite{Li:Yuan:Wu:2024}.

Our observations with Swift/XRT were also triggered while \J{} was active, Assuming a DM-inferred distance of 1.4--3~kpc, we calculate an upper limit on the 0.3--10~keV luminosity of $5.2\times10^{31}{\rm\,ergs\,s^{-1}}$ or $1.3\times10^{32}{\rm\,ergs\,s^{-1}}$ assuming a blackbody spectrum \citep[$kT = 0.3{\rm\,keV}$, representative of a magnetar;][]{Hurley-Walker:Zhang:Bahramian:2022} or a power law spectrum \citep[$\Gamma = 1$, representative of a polar;][]{Rodriguez+2023} specified above, respectively. This is about a factor of 6 lower than DART/ASKAP~J1832$-$0911, suggesting different origins.

Magnetars often have radiative outbursts accompanying radio emission of X-ray bolometric luminosities of $L_X \gtrsim 10^{34}{\rm\,ergs\,s^{-1}}$ \citep{CotiZelati+2018}, two orders of magnitude higher than our limit. However, \J{} could be a quiescent magnetar, though it would be among the least luminous quiescent magnetars known \citep{CotiZelati+2018, Hu+2019}. \red{Alternatively, \J{} could be a high magnetic field rotationally powered NS. For example, PSR~1937$-$3333 has an $L_{X} \approx 2\times 10^{32}{\rm\,ergs\,s^{-1}}$ \citep{olausen:zhu:2013}, which is at the edge of our detection limit.}

If \J{} contains a magnetic WD, it could be an intermediate polar system where the spins are synchronized with the orbital period. While typical intermediate polars have luminosities $L_X > 10^{33}{\rm\,ergs\,s^{-1}}$ \citep{Salcedo+2024}, \J{} could belong to the class of low-luminosity intermediate polars where $L_X \approx 10^{30-31}{\rm\,ergs\,s^{-1}}$, for which only a handful are known \citep{Salcedo+2024}.

\red{We conclude from {\it Swift XRT} observations that we can rule out the brightest X-ray emitters such as active magnetars. Still, we cannot rule out high magnetic field NSs or low-luminosity intermediate polars. Though the latter are likely rarer}


 \subsection{Negative Period Derivative}
Through timing, we find that the period derivative is nonphysical (negative at $>80\sigma$) for an isolated source. \red{This suggests additional intriguing physical mechanisms of the system, such as accreting matter from a companion,} which we will discuss in this section. We note that \J{} is the first LPT to \red{exhibit a significant negative $\dot{P}$.} GLEAM-X~J0704$-$37 \citep{Hurley-Walker:2024} have also shown hints of negative period derivatives, albeit at low significance, and likely caused by noise fluctuations of the time of arrivals. A similar scenario for the spin-up of \J{} could be that infalling material from a binary companion is providing \J{} with angular momentum. \red{Transitional millisecond pulsars \citep{Wijnands:1998,papitto:2022} and cataclysmic variables \citep{1995cvs..book.....W,Screiber:Belloni:Gansicke:2021} have been observed to have spin ups ranging from $\dot{\nu}\approx4-8\times10^{-13}$Hz/s \citep[see section 6.3.2 of][]{papitto:2022} and $\dot\nu\approx10^{-7}-10^{-12}$Hz/s \citep[see Table 1 of][]{Schaefer:2024}, respectively. These measurements have furthered the understanding of millisecond pulsars and the evolution of binary white dwarves. Therefore, if the radio period of \J{} was only associated with the central engine, then accretion is a likely scenario.}

\cite{Rodriguez:2025} found that the orbital period of GLEAM-X~J0704$-$37  was equal to the radio pulse period of 2.9 hours \citep{Rodriguez:2025}. \ILT{} is also a similar system \citep{deRuiter:2024}. If we speculate that \J{} is also in a compact binary where the radio pulse period and the orbital period are locked at 841\,s (14.02\,minutes), then a natural explanation for the significant $\dot{P}$ is orbital decay via gravitational wave emission. Indeed, such systems have already been observed in short-period WD-WD binaries \citep{burdge:2019,burdge:fuller:2019}, with orbital periods of 6.91 and 20.6 minutes.

We calculate the chirp mass, $M_c$, given our timing constraints by solving for $M_c$ in the equation
\begin{equation}
    2\dot{f} = \frac{96}{5}\pi^{(8/3)}(2f)^{(11/3)}\frac{(GM_c)^{(5/3)}}{c^5}
    \label{eq:chirp}
\end{equation}
where $\dot{f}$ is the orbital spin-up rate, $f$ is the orbital frequency, $G$ is the gravitational constant, and $c$ is the speed of light. We find that the chirp mass is $M_c = 0.36$ solar masses and defined by
\begin{equation}
    M_c=\frac{(m_1m_2)^{3/5}}{(m_1+m_2)^{1/5}}.
\end{equation}
By integrating Equation \ref{eq:chirp}, we also calculate that the merger time scale is $\sim$1,100,000 years, and assuming a circular orbit, Kepler's third law gives us that the semi-major axis is 0.00065\,AU (15 Earth Radii, 0.14 Sun Radii).

This constrains the component stars to WD-NS or WD-WD systems. In the WD-WD scenario, both stars must be at the upper range of WD masses. \red{Known cases of main sequence stars binaries which are compact enough, such as AM Canum Venaticorum, have component stars which are only $\sim 0.1$ solar masses \citep{Solheim:2010}, too low mass to work in this scenario. We further note that AM Canum Venaticorum systems are predicted to produce detectable gravitational waves by future space-based observatories \citep{Solheim:2010}. Given what we speculate here, \J{} could be an additional candidate for such observatories.}

This analysis assumes that the orbit is shrinking purely due to angular momentum loss via gravitational wave (GW) radiation. However, other factors may also affect the orbit, such as tidal deformation and magnetic field interactions between the component stars \citep{burdge:2019}.



\subsection{4206\,s bursting pattern}
\red{We showed in Section \ref{sec:timing} and Appendix \ref{sec:period_ambiguity} that given a 841\,s period, we see an over abundance of pulses which agree with a 4206\,s period. The pulsing pattern of \J{} is unique. Usually, one pulse is produced every five rotations of the 841\,s period (i.e., one pulse every 4206\,s period), but occasionally, two pulses can be produced every five rotations. When this happens, it is nearly always offset by 2 rotations of the 841\,s (or 0.4 in phase of the 4206\,s period). }Pulsars and LPTs are known to produce interpulses \citep{Manchester:lyne:1977,lee:caleb:2025}; therefore, we may be detecting a main pulse and an interpulse.   For the interpulse scenario to be accurate, we must assume four interpulse components, all separated by almost exactly a phase of 0.2. Heretofore, no pulsar or WD system has been observed to exhibit such an interpulse structure. Therefore, we believe that the interpulse scenario is unlikely.

\red{An alternative hypothesis is that \J{} is in a binary system with spin-orbit resonances. Indeed, due to this effect, Bloot et al. (2025) propose a 2103\,s binary orbit in a 5:2 resonance. Although we find no evidence for the specific configuration proposed by Bloot et al. (2025), the discussion presented above argues that a binary model may be necessary to explain the observations.}

\subsection{Is \J{} an NS or WD?}
 \paragraph{Neutron Star Model:}
    \red{As previously discussed, NSs have been observed to emit pulsed coherent radio emission as luminous or even more so in the form of FRBs and giant pulses, while being fully circularly polarized.}
    While our X-ray non-detection in the active state of \J{} can rule out active magnetars, they do not preclude rotationally-powered X-ray emitting NSs.
    The microstructure of \J{} is strongly reminiscent of other confirmed pulsars like XTE~1810$-$197 \red{\citep{maan:joshi:surnis:2019}} and PSR~J0901$-$4046 \citep{Caleb:2022}. \red{No such microstructure with significant quasiperiod has been identified in pulsating WDs before.}
    Finally, \J{} is likely in a binary system. Neutron star systems have shown they can exist in binaries and exhibit spin-up. For example, in the case of transitional millisecond pulsars \citep{Amruta:Hesssels:Archibald:2018,falanga:2005}.
\paragraph{White Dwarf Model}
    \red{There is an optical counterpart at the edge of the 3-$\sigma$ localization of \J{} which could indicate a 20000-40000\,K WD. However, we estimate a chance alignment probability of $\approx$6\%. Deeper observations are required.}

    The WD binary luminosities pose a problem for \J{}. From the dispersion measure, \J{} resides at a distance of 1.4--3\,kpc, with a peak flux of $\sim$0.4-9\,Jy. The two WD pulsars, \ARSCO{} \citep{Marsh:Gansicke:hummerich:2016,Buckley:2017} and \pelisoliWD{} \citep{Pelisoli:2023}, were found at a distance of 117\,pc and 237\,pc, respectively with flux densities of $\sim$ 4-10\,mJy. While the total fluence of their pulses is not provided, if we assume similar effective widths, the energies emitted by \J{} are 5-6 orders of magnitude greater than those emitted by \ARSCO{} and \pelisoliWD{}. Therefore, we conclude that the magnetic interaction mechanisms proposed for the WD pulsars will struggle to explain the emissions by \J{}.
    Compared to WD-MD binaries, \ILT{} and \GLEAM{}, \J{} emits radio bursts 2-3 orders of magnitude greater in energy. Therefore, we conclude that similar emission mechanisms are unlikely.
    
    Finally, like pulsars, WD binaries can be in cataclysmic variables which exhibit spin-up, for example, in intermediate polar systems \citep[e.g.][]{paice:scaringi:segura:2024}.
    
Given \red{these arguments, particularly those regarding the luminosity and polarization of \J{}, we conclude \J{} is more likely to be a NS-like system than a WD-like system}.

To conclude, we have discovered a fully circularly polarized long period radio transient (LPT) source, CHIME J1634+44. It has been localized with the Karl Jansky Very Large Array to $\sim1$\arcsec{} precision. CHIME J1634+44 possesses a period of 841\,s ($\sim$14 minutes) but clearly shows a secondary period at 4206\,s, possibly associated with binary activity. Furthermore, CHIME J1634+44 has a significantly negative period derivative, implying intriguing physics, such as gravitational wave radiation, which could be occurring. From arguments due to luminosity and polarization, we believe that \J{} is NS-like. \J{} will serve as an important test bed for LPT emission theories and is unique among the array of known transient source emitters.

\section{Acknowledgments}
We thank Ryan Lynch for his support in calibrating the GBT data.
\allacks{}

\section{Software and third party data repository citations} \label{sec:cite}

%
\vspace{5mm}
\facilities{CHIME, GBT, VLA, Swift(XRT and UVOT), HSC}


\software{Numpy, Astropy, sigpyproc3, Presto}



\appendix
\section{Detections}
\label{sec:appendix}
We provide tables for all the detections from CHIME/FRB, CHIME/Pulsar, the GBT, and the VLA in Tables \ref{tab:appen_tab1}, \ref{tab:appen_tab2}, and \ref{tab:appen_tab3}.
Where possible, we provide the fluence, effective width, peak flux density, and pseudo-luminosity. The MJD is referenced at 800\,MHz for CHIME/Pulsar, 400\,MHz for CHIME/FRB, 920\,MHz for the GBT and 1440\,MHz for the VLA.
\begin{table}
  \caption{Detection results for CHIME/Pulsar observations}
  \hspace{-2cm}
  \begin{tabular}{lllllll}
    \hline
    \hline
    Observation  & MJD (Topocentric) & Fluence (Jyms) & Effective width (ms) & Peak flux density (Jy) & Pseudo-luminosity$^{*}$ (Jykpc$^{2}$) \\
    \hline
    \hline
    CHIME/Pulsar & 60270.84871(2)    & 140(30)        & 80(30)               & 1.7(3)                 & 3.3                                   \\
    CHIME/Pulsar & 60301.76199(1)    & 900(200)       & 300(100)             & 3.2(6)                 & 6.3                                   \\
    CHIME/Pulsar & 60303.75793(2)    & 2000(400)      & 500(200)             & 3.9(8)                 & 7.6                                   \\
    CHIME/Pulsar & 60305.753885(9)   & 1000(200)      & 300(100)             & 3.4(7)                 & 6.7                                   \\
    CHIME/Pulsar & 60307.749838(8)   & 2100(400)      & 700(300)             & 2.8(6)                 & 5.5                                   \\
    CHIME/Pulsar & 60309.74578(1)    & 1100(200)      & 500(200)             & 2.3(5)                 & 4.5                                   \\
    CHIME/Pulsar & 60311.74173(2)    & 1400(300)      & 700(300)             & 2.1(4)                 & 4.1                                   \\
    CHIME/Pulsar & 60315.73362(2)    & 5000(1000)     & 700(300)             & 8(2)                   & 15.7                                  \\
    CHIME/Pulsar & 60317.729565(7)   & 2000(400)      & 400(200)             & 4.7(9)                 & 9.2                                   \\
    CHIME/Pulsar & 60336.66665(2)    & 4400(900)      & 400(100)             & 12(2)                  & 23.5                                  \\
    CHIME/Pulsar & 60338.66260(1)    & 700(100)       & 230(90)              & 3.2(6)                 & 6.3                                   \\
    CHIME/Pulsar & 60340.658534(7)   & 500(100)       & 220(90)              & 2.3(5)                 & 4.5                                   \\
    CHIME/Pulsar & 60342.65446(2)    & 2000(400)      & 500(200)             & 3.9(8)                 & 7.6                                   \\
    CHIME/Pulsar & 60344.65040(2)    & 1700(300)      & 600(300)             & 2.6(5)                 & 5.1                                   \\
    CHIME/Pulsar & 60346.646318(9)   & 2000(400)      & 700(300)             & 2.9(6)                 & 5.7                                   \\
    CHIME/Pulsar & 60348.64226(1)    & 1500(300)      & 300(100)             & 4.4(9)                 & 8.6                                   \\
    CHIME/Pulsar & 60349.63525(2)    & 1400(300)      & 700(300)             & 1.8(4)                 & 3.5                                   \\
    CHIME/Pulsar & 60350.63819(2)    & 1600(300)      & 400(200)             & 3.9(8)                 & 7.6                                   \\
    CHIME/Pulsar & 60351.63116(1)    & 270(50)        & 150(60)              & 1.8(4)                 & 3.5                                   \\
    CHIME/Pulsar & 60352.63411(1)    & 4000(800)      & 800(300)             & 5(1)                   & 9.8                                   \\
    CHIME/Pulsar & 60353.62709(1)    & 800(200)       & 190(80)              & 4.0(8)                 & 7.8                                   \\
    CHIME/Pulsar & 60357.61895(2)    & 700(100)       & 300(100)             & 2.6(5)                 & 5.1                                   \\
    CHIME/Pulsar & 60359.61488(1)    & 1000(200)      & 170(70)              & 6(1)                   & 11.8                                  \\
    CHIME/Pulsar & 60375.562934(6)   & 60(10)         & 70(30)               & 0.9(2)                 & 1.8                                   \\
    CHIME/Pulsar & 60377.55887(1)    & 2400(500)      & 500(200)             & 4.5(9)                 & 8.8                                   \\
    CHIME/Pulsar & 60379.55481(2)    & 1200(200)      & 500(200)             & 2.5(5)                 & 4.9                                   \\
    CHIME/Pulsar & 60383.54667(1)    & 2100(400)      & 400(100)             & 6(1)                   & 11.8                                  \\
    CHIME/Pulsar & 60384.539665(4)   & 80(20)         & 160(60)              & 0.5(1)                 & 1.0                                   \\
    CHIME/Pulsar & 60387.538556(9)   & 460(90)        & 220(90)              & 2.1(4)                 & 4.1                                   \\
    CHIME/Pulsar & 60388.53154(2)    & 70(10)         & 180(70)              & 0.42(8)                & 0.8                                   \\
    CHIME/Pulsar & 60390.52750(1)    & 500(100)       & 140(50)              & 3.5(7)                 & 6.9                                   \\
    CHIME/Pulsar & 60392.523456(9)   & 1800(400)      & 500(200)             & 3.8(8)                 & 7.4                                   \\
    CHIME/Pulsar & 60394.519381(7)   & 1700(300)      & 300(100)             & 6(1)                   & 11.8                                  \\
    CHIME/Pulsar & 60404.47981(1)    & 2400(500)      & 300(100)             & 9(2)                   & 17.6                                  \\
    CHIME/Pulsar & 60412.46367(2)    & 1700(300)      & 700(300)             & 2.7(5)                 & 5.3                                   \\
    CHIME/Pulsar & 60414.45965(3)    & 900(200)       & 700(300)             & 1.3(3)                 & 2.5                                   \\
    CHIME/Pulsar & 60417.448649(7)   & 800(200)       & 300(100)             & 2.7(5)                 & 5.3                                   \\
    CHIME/Pulsar & 60421.44062(1)    & 1300(300)      & 400(200)             & 3.4(7)                 & 6.7                                   \\
    CHIME/Pulsar & 60423.43660(2)    & 1500(300)      & 500(200)             & 2.8(6)                 & 5.5                                   \\
    CHIME/Pulsar & 60458.34241(1)    & 150(30)        & 180(70)              & 0.8(2)                 & 1.6                                   \\
    CHIME/Pulsar & 60481.272590(7)   & 100(20)        & 120(50)              & 0.8(2)                 & 1.6                                   \\
    CHIME/Pulsar & 60516.179617(7)   & 900(200)       & 300(100)             & 2.7(5)                 & 5.3                                   \\
    CHIME/Pulsar & 60518.175765(4)   & 50(10)         & 110(40)              & 0.47(9)                & 0.9                                   \\
    CHIME/Pulsar & 60522.15821(1)    & 500(100)       & 150(60)              & 3.5(7)                 & 6.9                                   \\
    \hline
    \hline
    \\
  \end{tabular}

  \label{tab:appen_tab1} $^{*}$ This is given as a lower limit by using the
  lower estimate of the distance, 1.4\,kpc.
\end{table}

\begin{table}
  \caption{Detection results for CHIME/FRB observations}
  \begin{tabular}{ll|ll}
    \hline
    \hline
    Observation        & MJD (Topocentric) & Observation         & MJD (Topocentric) \\
    \hline
    \hline
    CHIME/FRB metadata & 58893.62626(1)    & CHIME/FRB intensity & 59883.914336(9)   \\
    CHIME/FRB metadata & 58955.45209(1)    & CHIME/FRB intensity & 60003.584379(9)   \\
    CHIME/FRB metadata & 58990.35777(1)    & CHIME/FRB intensity & 60270.848702(8)   \\
    CHIME/FRB metadata & 58994.34984(1)    & CHIME/FRB intensity & 60274.84070(1)    \\
    CHIME/FRB metadata & 59021.27218(1)    & CHIME/FRB intensity & 60276.836688(6)   \\
    CHIME/FRB metadata & 59023.26824(1)    & CHIME/FRB intensity & 60278.83267(2)    \\
    CHIME/FRB metadata & 59025.26430(1)    & CHIME/FRB intensity & 60280.82866(1)    \\
    CHIME/FRB metadata & 59027.26036(1)    & CHIME/FRB intensity & 60303.757945(5)   \\
    CHIME/FRB metadata & 59052.18699(1)    & CHIME/FRB intensity & 60305.753897(9)   \\
    CHIME/FRB metadata & 59056.17917(1)    & CHIME/FRB intensity & 60307.749847(9)   \\
    CHIME/FRB metadata & 59124.99057(1)    & CHIME/FRB intensity & 60309.74579(1)    \\
    CHIME/FRB metadata & 59241.67340(1)    & CHIME/FRB intensity & 60311.74174(1)    \\
    CHIME/FRB metadata & 59276.57788(1)    & CHIME/FRB intensity & 60313.73769(1)    \\
    CHIME/FRB metadata & 59452.09220(1)    & CHIME/FRB intensity & 60315.733622(9)   \\
    CHIME/FRB metadata & 59599.69042(1)    & CHIME/FRB intensity & 60338.66260(1)    \\
    CHIME/FRB metadata & 59601.68636(1)    & CHIME/FRB intensity & 60340.658543(7)   \\
    CHIME/FRB metadata & 59630.60303(1)    & CHIME/FRB intensity & 60342.65447(2)    \\
    CHIME/FRB metadata & 59634.59489(1)    & CHIME/FRB intensity & 60344.65040(1)    \\
    CHIME/FRB metadata & 59636.59083(1)    & CHIME/FRB intensity & 60346.64633(2)    \\
    CHIME/FRB metadata & 59638.58675(1)    & CHIME/FRB intensity & 60348.64227(2)    \\
    CHIME/FRB metadata & 59665.50760(1)    & CHIME/FRB intensity & 60349.63526(2)    \\
    CHIME/FRB metadata & 59667.50357(1)    & CHIME/FRB intensity & 60350.63820(1)    \\
    CHIME/FRB metadata & 59669.49952(1)    & CHIME/FRB intensity & 60351.63117(1)    \\
    CHIME/FRB metadata & 59700.41284(1)    & CHIME/FRB intensity & 60353.627102(9)   \\
    CHIME/FRB metadata & 59702.40883(1)    & CHIME/FRB intensity & 60357.618958(5)   \\
    CHIME/FRB metadata & 59707.39385(1)    & CHIME/FRB intensity & 60377.55888(1)    \\
    CHIME/FRB metadata & 59709.38986(1)    & CHIME/FRB intensity & 60379.55482(2)    \\
    CHIME/FRB metadata & 59733.32280(1)    & CHIME/FRB intensity & 60390.52751(1)    \\
    CHIME/FRB metadata & 59737.31488(1)    & CHIME/FRB intensity & 60412.46368(1)    \\
    CHIME/FRB metadata & 59739.31093(1)    & CHIME/FRB intensity & 60414.45966(2)    \\
    CHIME/FRB metadata & 59742.30003(1)    & CHIME/FRB intensity & 60421.44063(1)    \\
    CHIME/FRB metadata & 59744.29607(1)    & CHIME/FRB intensity & 60423.43661(2)    \\
    CHIME/FRB metadata & 59746.29213(1)    & CHIME/FRB intensity & 60458.342412(8)   \\
    CHIME/FRB metadata & 59777.20691(1)    &                     &                   \\
    CHIME/FRB metadata & 59806.12589(1)    &                     &                   \\
    CHIME/FRB metadata & 60182.09721(1)    &                     &                   \\
    \hline
    \hline
  \end{tabular}

  \label{tab:appen_tab2}
\end{table}

\begin{table}
  \caption{Detection results for VLA and GBT observations}
  \hspace{-2cm}
  \begin{tabular}{llllll}
    \hline
    \hline
    Observation & MJD (Topocentric) & Fluence (Jyms) & Effective width (ms) & Peak flux density (Jy) & Pseudo-luminosity$^{*}$ (Jykpc$^{2}$) \\
    \hline
    \hline
    VLA         & 60301.76197(1)    & --             & --                   & --                     & --                                    \\
    VLA         & 60301.85935(1)    & --             & --                   & --                     & --                                    \\
    GBT         & 60311.54701(1)    & 3100(300)      & 900(90)              & 3.5(4)                 & 6.9                                   \\
    GBT         & 60311.59569(2)    & 2900(300)      & 1100(100)            & 2.7(3)                 & 5.3                                   \\
    \hline
    \hline
    \\
  \end{tabular}
  $^{*}$ This is given as a lower limit by using the lower estimate of the distance,
  1.4\,kpc.

  \label{tab:appen_tab3}
\end{table}

\subsection{Flux Density and Fluence}\label{sec:flux}
Flux density is difficult to estimate for CHIME/FRB due to a complex beam response \citep{andersen2023}. Furthermore, as the CHIME/FRB total intensity data segment can be slightly shorter than the pulse width, it is difficult to get a steady off-source measurement. For this reason, we provide CHIME/FRB data only in arbitrary units. In general, flux calibration with the CHIME/Pulsar instrument is simpler due to the tracking of phased array beams and longer observation tracks. With the CHIME/Pulsar instrument, daily calibration observations are taken. We use three calibrators that bracket the \J{} declination range. These calibrators are 3C380, 3C196, and NGC1265 at declinations of +48.75$^\circ$, +48.22$^\circ$, and +41.9$^\circ$, respectively. The calibrators are used to determine the telescope temperature, $T_{\rm telescope}$, which includes the receiver, structure, and ground temperatures. The calibration is done by solving the following equation:
\begin{equation}
    S_{\rm src}=\frac{(T_{\rm sky}+T_{\rm telescope})\times ({\rm ON}-{\rm OFF})}{G\times {\rm OFF}}, 
\end{equation}
where $S_{\rm src}$ is the flux density of the calibrator in Jy, ${\rm ON}$ and ${\rm OFF}$ are the raw counts of the on and off source integrations respectively, and $G$ is the gain of the phased array beam in K/Jy. $T_{\rm sky}$ is the sky temperature in K, which is determined by the Haslam 408\,MHz all-sky continuum map \citep{haslam:1982}. $T_{telescope}$ is calibrated by fitting a polynomial to the catalog values of the calibrators. A full discussion of the CHIME/Pulsar flux density calibration pipeline is given by \cite{Dong_2024}. The fluence is the time-integrated value across the whole burst, and the effective width is the fluence divided by the peak flux density.
\subsection{Dispersion Measure}
\label{sec:discovery_dm}
\begin{figure}
    \centering
    \includegraphics[width=0.5\linewidth]{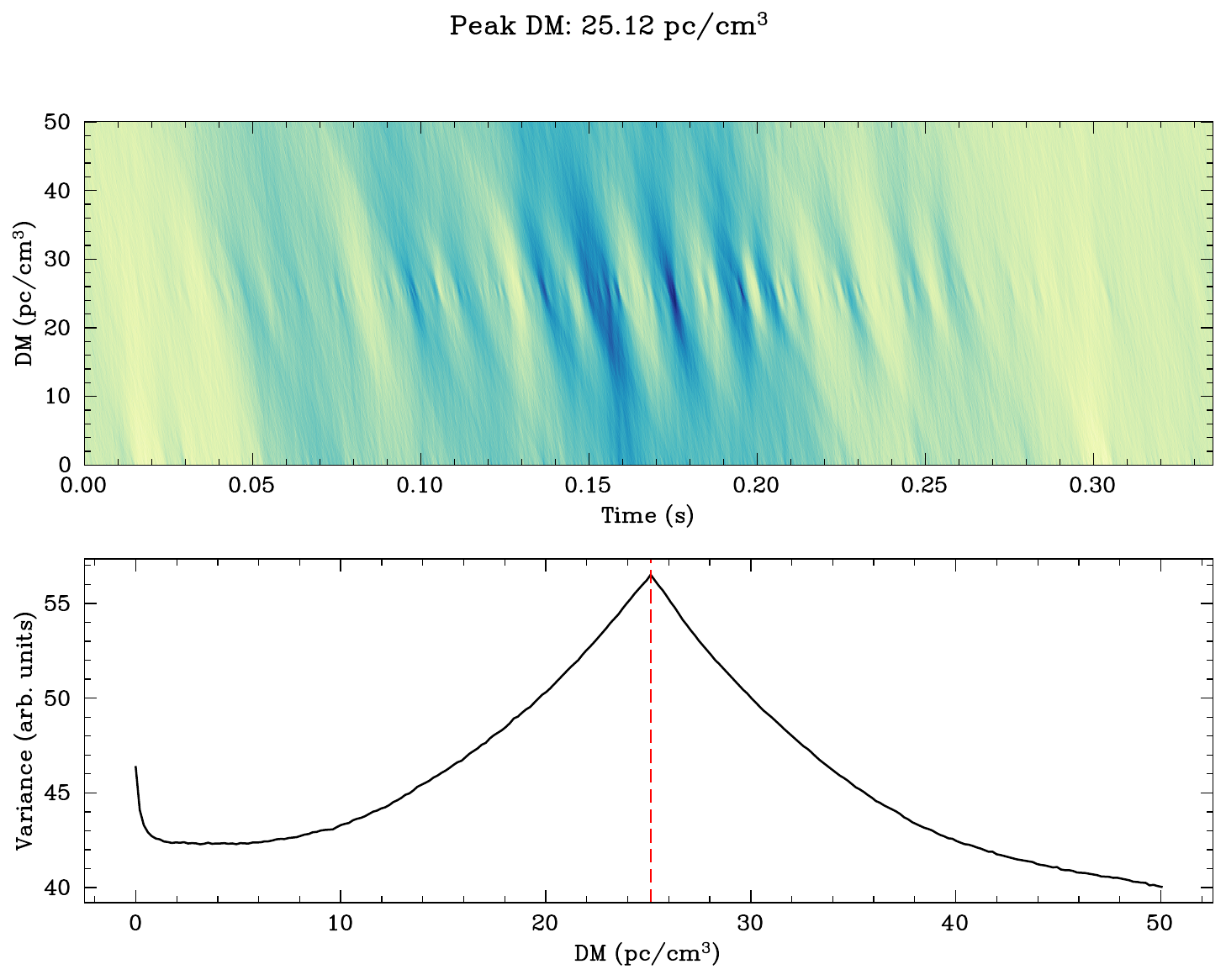}
    \caption{Example of a DM-time power spectrum from burst 60311A, detected by CHIME/Pulsar between 400-800\,MHz. The bottom panel shows the variance across the DM of the top panel.}
    \label{fig:dm_time}
\end{figure}

The DM is determined by calculating the variance over time in the dispersion-time power spectrum for each trial dedispersion (lowest panel in Figure \ref{fig:example_det}).
The variance is used instead of the summed intensity because as the hot-spot structure spreads, the total power is distributed over time but not significantly diminished. We then find the peak that is not at 0, which is taken to be the optimal DM of the burst. We show an example in Figure \ref{fig:dm_time}. The mean of the DM for all bursts is calculated to be 25.0(2)~pc\,cm$^{-3}$, where the uncertainty is the standard error on the mean. Throughout this work, we assume that this is the true DM of \J{}. We note that there appears to be no significant DM evolution of \J{}.

The inferred DM distance for this direction is 1.4\,kpc and 3.0\,kpc for the NE2001 \citep{Cordes:2001} and YMW16 \citep{Yao:Manchester:2016} dispersion models, respectively.

\subsection{Time of Arrival Determination}
Arrival times for pulses are required for determining a phase-coherent timing model \J{}. Because of significant microstructure in the pulses, we extract this information by applying a 200\,ms Gaussian filter across each burst so that the microstructure is fully smoothed over. This technique has similarly been used by \cite{Hurley-Walker:Zhang:Bahramian:2022} and \cite{Dong:2024}. The TOA and 1~$\sigma$ errors are defined as the peak and the full-width-half-maximum of the smoothed profile, respectively.
For the metadata-only detections, we report these bursts with a TOA error of 1\,s and assume that the burst has been dedispersed to the optimal DM by the CHIME/FRB Bonsai pipeline \citep{10.3847/1538-4357/aad188}.

\subsection{CHIME Polarization Measurements}\label{sec:pol_chime}

From August 2023 through April 2024, 24 observed bursts from \J{} triggered the raw voltage data recording system for CHIME/FRB \citep{Michilli+2021, basecat1}.
While the overarching burst envelopes have durations of $\sim$seconds (Figure~\ref{fig:example_det}), due to data buffer constraints, only $\sim$100~ms of raw voltage data are saved per burst.
These voltage data are saved with a 2.56~$\mu$s time and $\sim$390~kHz frequency resolution, and full Stokes information.
We can thus conduct polarimetry analysis on these bursts despite the limited temporal extent of the written data.

We use the CHIME/FRB polarimetry pipeline developed for FRBs, as detailed by \citet{Mckinven+2021} and \citet{Pandhi+2024}, with a few modifications to the defaults described by those works.
Off-burst data are necessary for determining noise statistics when obtaining polarization fractions. While the standard voltage data products for FRBs encompass both on-burst and off-burst extents, for \J{}, there is no true off-burst data within the data span.
To remedy this, for the seven events where sub-burst structure in the raw voltage data were S/N~$>$~10, we take their respective written data and beamform the files to both the best position provided by VLA/\textit{realfast} for our on-burst data and 0.5~degrees lower in declination for our off-burst data.
We form our off-position beam lower in declination, as the CHIME primary beam model varies much more rapidly in RA than in Dec \citep{10.3847/1538-4365/ac6fd9}.
All bursts were de-dispersed to $\mathrm{DM} = 25.0$~pc~cm$^{-3}$.

The Faraday rotation measure (RM) is determined by both non-parametric methods \citep[rotation measure synthesis;][]{1966MNRAS.133...67B, 2005A&A...441.1217B} and parametric methods \citep[QU-fitting; adapted from the \texttt{RM-tools} package,][]{2020ascl.soft05003P}.
Both measurements suffer from instrumental systematics, particularly a non-trivial leakage effect between Stokes $U$ and $V$ values introduced by a non-zero physical delay between the two linear polarizations (cable delay).
This cable delay can also introduce a sign ambiguity to the measured RM signal.
Rather than adopt the default model used by \citet{Mckinven+2021} and \citet{Pandhi+2024} where the assumption is that there is no significant circular polarization intrinsic to the burst (i.e., $|V|/I \lesssim 0.2$),
we adopt a model where the QU-fitting routine allows for a non-zero circularly polarized component with a power-law spectrum.

\J{} appears to have a low RM, with detections of signal ranging from $\mathrm{RM} \sim -12$~rad~m$^{-2}$ to $\sim +11$~rad~m$^{-2}$, with no clear sign preference between the RM synthesis and QU-fitting results.
We check the Galactic RM map by \citet{hutschenreuter+2022_MW_RM_map} and find $\mathrm{RM_{gal}} = 19 \pm 5$~rad~m$^{-2}$; this measurement is a cumulative sum of RM along the entire Galactic line of sight.
A nearby pulsar $\sim$5$^\circ$ away, J1638+4005, also has $\mathrm{RM} = 17$~rad~m$^{-2}$.
We thus conclude that \J{} has an RM in line with expectations from the intervening interstellar medium, and that there is no evidence for a dense, magnetized circumburst environment.
Furthermore, our measurements are consistent with Bloot et al. 2025.

Consistent with the VLA/\textit{realfast} data, all these CHIME bursts showed evidence for significant intrinsic circular polarization, with $|V|/I \gtrsim 0.9$.
All bursts also showed evidence for non-negligible intrinsic linear polarization, with $|L|/I \gtrsim 0.2$.
Exact fractions and uncertainties are not reported, as despite the attempts to correct for instrumental effects, there are likely still instrumental sources of circular polarization of $\gtrsim$20\% \citep{Pandhi+2024}.

\subsection{Karl G.\ Jansky Very Large Array}\label{sec:vla}
\label{subsec:discovery_VLA}
\begin{figure}
\centering
\includegraphics[width=0.3\textwidth]{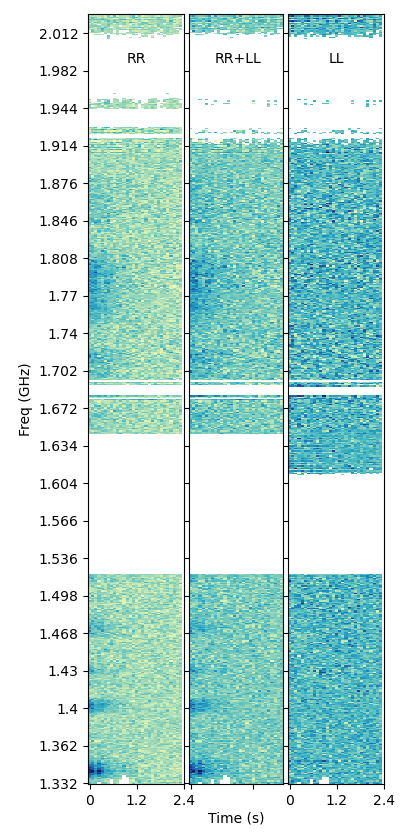}
\caption{Dynamic spectrum for a portion of the first burst of \J{} seen by VLA/\textit{realfast}. The realfast search system detected this burst at the boundary of a buffer, so this plot only shows the second half of the first burst. The three panels show the RR, RR+LL (Stokes I), and LL polarization products. The time and frequency span of all three panels is identical.}
\label{fig:vlaburst}
\end{figure}

To better localize the radio bursts and search for multiwavelength counterparts, we observed \J{} with the Karl G. Jansky Very Large Array (VLA) under program code 23B-337. The observations used the \textit{realfast} commensal fast transient search system to detect the bursts, \red{which outputs 10\,ms visibilities, much shorter than the shortest 2\,s integrations of standard VLA observations} \citep{realfast}.

A single 4-hour observation was executed on 23 December 2023. The VLA was in its most compact (``D'') configuration with 26 antennas available. We observed the position of \J{} in the 1--2 GHz (``L'') band, which is the lowest supported by \textit{realfast}, and which minimized the extrapolation from the expected flux from CHIME. The synthesized beam size had a full-width at half-maximum of 46\arcsec. 

The observation was designed to provide basic gain calibration for the transient search. The flux calibrator was 3C286 and the gain calibrator was J1625+4134. Target scans were 2 minutes long and the on-target observing efficiency was roughly 85\%.

Each scan produced standard correlated visibility output with 2-second time resolution and a fast copy with 10-ms time resolution. The fast copy was sent to the \textit{realfast} transient search pipeline. The standard visibility data included all four cross-correlation products, but \textit{realfast} receives only the self-correlation products, RR and LL. Both data streams included a frequency span from 1--2 GHz, covered by 16 subbands of 64 channels each (channel width of 1 MHz).

The \textit{realfast} system forms Stokes I images over a range of DMs and timescales \citep{realfast}. The mean visibility is subtracted during the transient search, so any image with a significant source will likely be a transient signal (either astrophysical or interference). A simple threshold is used to trigger the recording of fast visibility data for offline analysis.

In total, this observation produced 17 candidates above a detection threshold of 7.2. The threshold was defined to allow 0.1 false events per scan, or roughly nine false events total. From these 17 candidates, two bursts were identified as astrophysical \red{with signal-to-noise ratios of 8.9 and 8.4 in images made on the longest real-time search width of 80 ms. The rest of the detections were RFI.}


Figure \ref{fig:vlaburst} shows the dynamic spectrum for the first burst. The triggered fast visibility data span 9 seconds, but only the \red{2.4} seconds near the burst are shown in the figure. The burst is seen across most of the observing band, but is strongest below 1450 MHz.

\subsection{Quasiperiodicity} \label{sec:quasiperiodicity}
\begin{figure}[ht!]
\centering
\includegraphics[width=\linewidth]{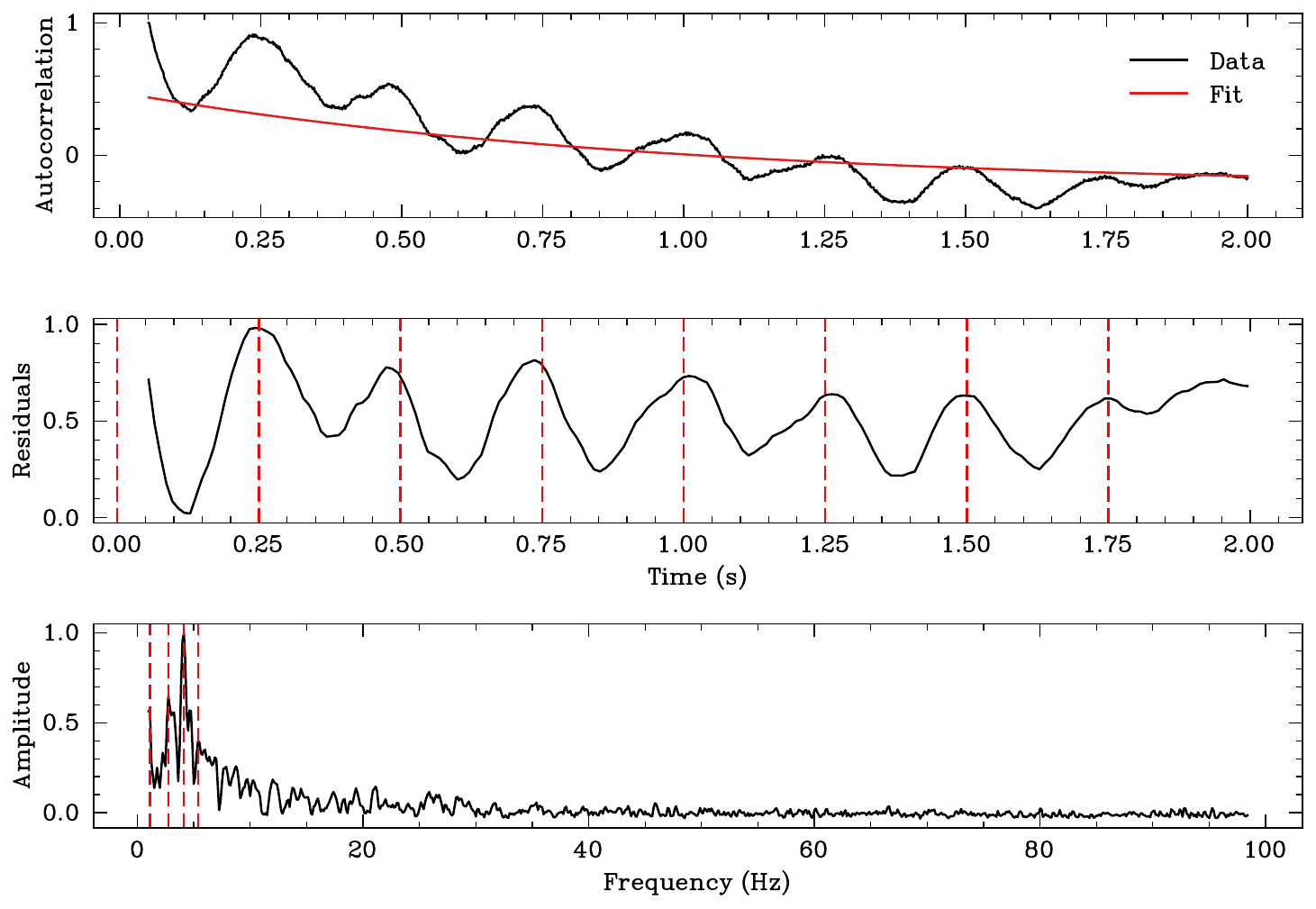}
\caption{An example of the autocorrelation function and the FFT power spectrum from burst 60309A. The top panel shows the raw autocorrelation function normalized to 1. The middle panel shows a subtraction of the exponentially decaying baseline, an approximation to the red noise. The quasiperiod is identified at $\sim$4~Hz and marked in the red dotted lines. The bottom panel shows the FFT power spectrum. The red dashed lines show the identified peaks.}
\label{fig:autocorr_example}
\end{figure}
\begin{figure}[ht!]
\centering
\includegraphics[width=0.49\linewidth]{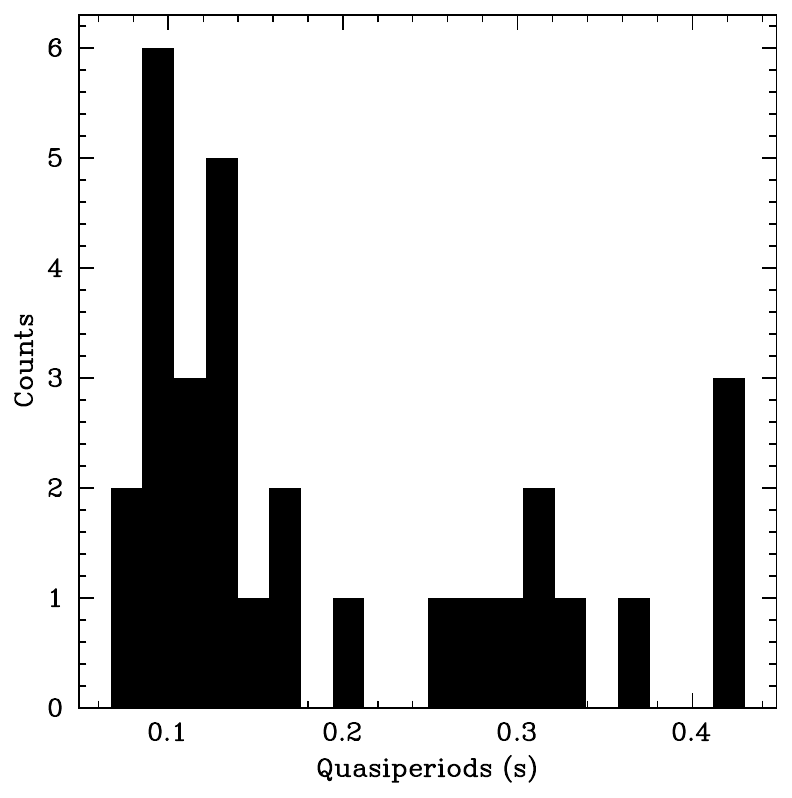}
\includegraphics[width=0.5\linewidth]{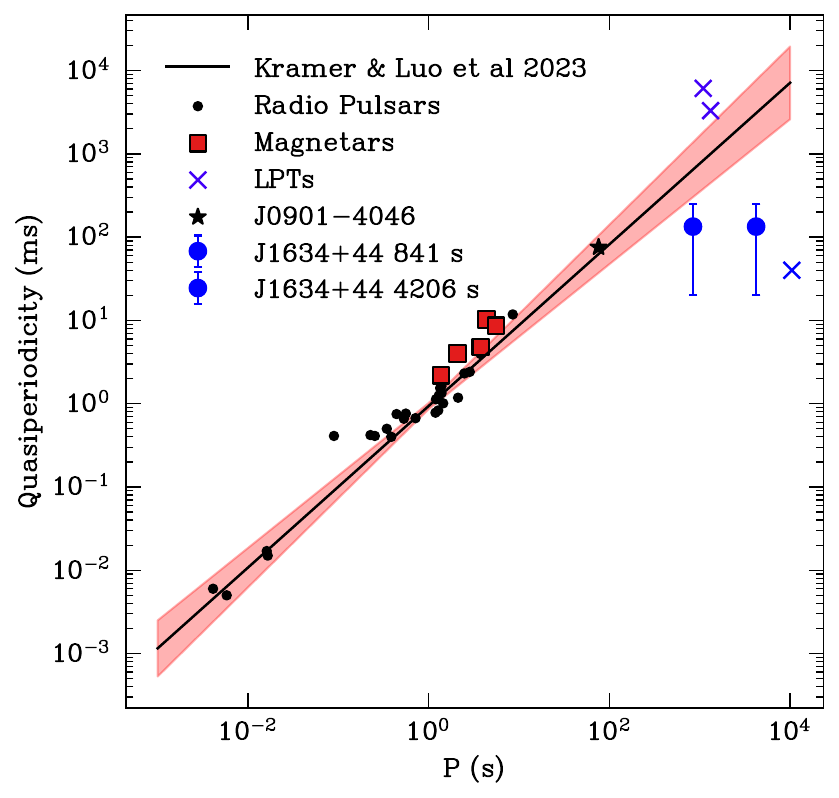}
\caption{Left panel: A histogram of all the quasiperiods measured for \J{}. Right panel: The relationship between quasiperiodicity and period. Due to the log scale, the errors at the rightmost of this figure are large. J1634+44 does not fall within the neutron star quasiperiod region.}
\label{fig:quasiperiod_relation}
\end{figure}

Quasiperiodicity is different from strict periodicity, such as that identified in Section \ref{sec:timing}, as they typically manifest as semi-cyclic peaks within a single burst as opposed to strict periodic separation between bursts. Quasiperiodicity is tested with an autocorrelation function where the time lag is calculated with respect to autocorrelation power. Specifically, we employ the \texttt{numpy.correlate} \citep{10.1038/s41586-020-2649-2} function on the dedispersed band-averaged time series of each burst. This is defined by
\begin{equation}
    c_k = \sum_{n} a_{n+k} \cdot \bar{a}_n,
\end{equation}
where $c_k$ is the correlation function and $a$ is the dedispersed band-averaged time series. $\bar{a}$ denotes complex conjugation, though our intensity time series is strictly real.

In addition to autocorrelation, we also perform a discrete Fast Fourier Transform (FFT) on each time series. This is done via the \texttt{sigpyproc}\footnote{\url{https://github.com/FRBs/sigpyproc3}} package. Only peaks greater than 3$\sigma$ are counted towards potential quasiperiodicity. Taking the largest peak identified in the FFT, we mark each timelag where we expect to see a peak in the autocorrelation. This is shown in Figure \ref{fig:autocorr_example}. This was verified manually, as sometimes a higher order harmonic of the actual quasiperiod may be mistakenly identified by the FFT. In all cases, only quasiperiodicities above 3~$\sigma$ in the FFT are considered. Quasiperiods are collated for all bursts and plotted in Figure \ref{fig:quasiperiod_relation}.

\cite{kramer:liu:2023} showed that almost all variants of isolated neutron stars fit a quasiperiod-period relation; this can include millisecond pulsars, slow pulsars, Rotating Radio Transients (RRATs) and magnetars. Indeed,PSR J0901--4046, an LPT with a spin period of 76~s, and assumed to be a magnetar \citep{Caleb:2024} sits well within the error region of the quasiperiod-period relation. The median quasiperiod for \J{} is $P_\mu=0.13(0.11)$~s. In Figure \ref{fig:quasiperiod_relation} we show \J{} amongst other NSs and LPTs with both the $\sim$4206~s and the $\sim$841~s period. This shows that \J{} is well outside the error region of the relation, especially if we assume the $\sim$4206~s period. This may hint that \J{} is unlikely to be an isolated neutron star.

\section{Period Ambiguity}\label{sec:period_ambiguity}
\subsection{Simulations}
\begin{figure}[ht]
\centering
\includegraphics[width=0.5\linewidth]{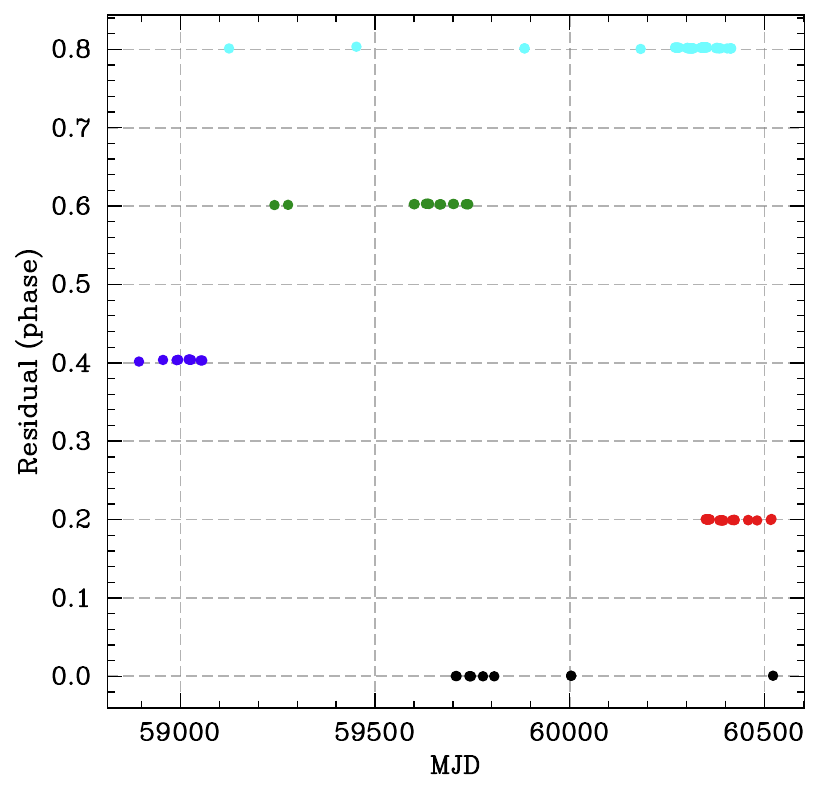}
\caption{TOA residuals for all the detections made given a 4206~s period, without any phase jumps included. \red{This shows that all bursts are arriving at a multiple of 841\,s, that is, bursts arrive at multiples of 0.2 in phase.} The colors represent the phase bins to which they belong.}
\label{fig:4206_timing_nojump}
\end{figure}

\begin{figure}
    \centering
    \includegraphics[width=0.49\linewidth]{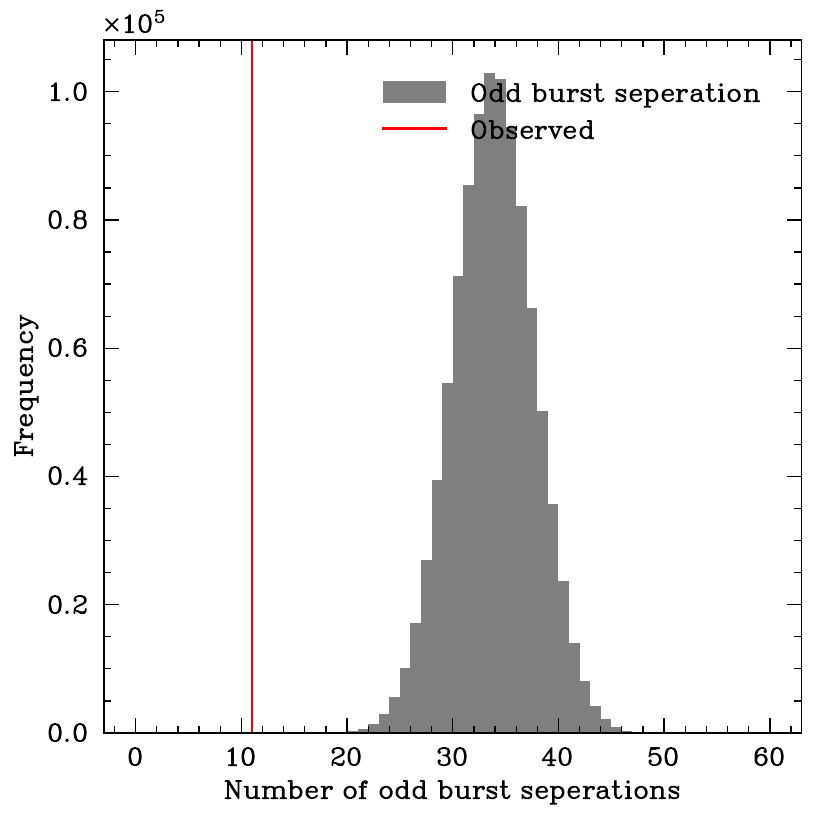}
    \caption{The distribution of the number of odd-day burst separations for 1,000,000 simulations of 60 burst pairs. The red line shows the observed number of odd burst separations.}
    \label{fig:simulations}
\end{figure}

We noticed that when \J{} enters an activity window, it mostly bursts every second day. \red{We define an activity window as when bursts are separated by, at most, 10 days. That is, if there were four bursts on day 1, 8, 13, and 100, there is an activity window from days 1--13 and a new one beginning on day 100. This is robust as bursts come in clusters where the intra-cluster separation is generally much less than 10 days and the inter-cluster separation is generally much larger.} As stated above, this repetition fits a beat frequency between 4206\,s and CHIME's daily observation cadence. The spin period of 4206\,s was also identified by Bloot et al. (2025). However, as shown in Figure \ref{fig:4206_timing_nojump}, taking all the data into account, it is clear that the bursts are not phase connected with a period of 4206\,s. Therefore, we asked the following question: ``If \J{} indeed possesses a period of 841\,s, how likely is it to see the burst pattern we observe with CHIME?''. This question is simplified by the fact that due to CHIME's short transit duration, if \J{} was persistently bursting with a 841\,s period, we would detect it once per day, every day. On the contrary, if \J{} were to persistently burst with a 4206\,s period, we should detect it every second day. To answer this question, we performed the following steps:
\begin{enumerate}
    \item For our observed bursts with CHIME, we define an arbitrary activity period of 10 days. That is, if bursts are detected within 10 days, we consider the bursts part of the same activity period. 
    \item We form consecutive pairs of bursts and count how many such pairs are detected to agree with {\it only} the 841\,s period (i.e., the pairs are separated by an odd number of days). We ignore all burst pairs that are not part of the same activity period. 
    \item We find that there are 60 burst pairs, 11 of which agree {\it only} with the 841\,s period, and 49 of which agree with both the 841\,s and 4206\,s period.
    \item We assume that if the period is truly 841\,s then we would expect to find the separation between bursts to be equally as likely to be an odd number of days or an even number of days. That is, at each 841\,s epoch, there is a 50\% chance a burst will occur.
    \item We simulate 1,000,000 iterations of 60 burst pairs and count the number of even and odd-day burst separations.
    \item We find that the mean number of odd-day burst separations is 33.35 with a standard deviation of 3.86. Therefore, our observations are $\sim5.8$ standard deviations away from the mean, with a probability of 3.5$\times10^{-9}$. This is shown in Figure \ref{fig:simulations}.
\end{enumerate}

To summarize, our simulations show that it is highly unlikely for \J{} to be an isolated ``normal'' 841\,s LPT and create the burst patterns that are observed. We conclude that additional, unknown physical mechanisms, such as orbital resonances, must be invoked to explain our observations. We note that Bloot et al. (2025) proposed a possible 5:2 spin-orbit resonance with an orbital period of $\sim$2103.1\,s. We find no evidence of this orbital period in our timing data.

\section{Swift X-ray and UV}\label{sec:xray}
Upon the detection of an activity period beginning in MJD 60270 (2023 November 22), we triggered 10~ks of {\it Swift} ToO time to observe \J{} on 2023 December 07 and 2024 January 04 (target ID 16412).
The data (XRT/Photon Counting mode) were processed with the \texttt{UK Swift Science Data Centre}\footnote{\url{https://www.swift.ac.uk/user_objects/}} online tool \citep{Evans+2020}.
In total, 6 sources were detected across the 0.3-10~keV band in the field of view of Swift/XRT; however, none were within a 3-$\sigma$ error region of \J.
We then used the upper limit server\footnote{\url{https://www.swift.ac.uk/LSXPS/ulserv.php}}, which utilizes the corresponding data (for a set of input coordinates), the background map, and the exposure map to determine an upper limit on the 0.2--10.0~keV count rate of $1.8\times10^{-3}{\rm\,c/s}$ \citep{Evans+2020}.
In order to derive physically meaningful comparisons, we utilized the \texttt{WebPIMMS} tool to calculate the flux upper limit for \J{}. We assumed $N_H=7.5\times 10^{20}$~cm$^{-2}$ (based on the source DM and the $N_H$-DM relation; \citealt{He+2013}), and either a blackbody spectrum with $kT=0.3$~keV or a power law spectrum with $\Gamma=1$. These values are representative of a magnetar \citep{Hurley-Walker:Zhang:Bahramian:2022} or an intermediate polar \citep[WD with MS companion;][]{Rodriguez+2023}. We found that with a 3-$\sigma$ upper limit on the 0.2--10.0~keV count rate of 1.8$\times 10^{-3}$~c/s, the corresponding 0.3--10~keV unabsorbed flux upper limit is $4.8\times10^{-14}$~ergs~cm$^{-2}$~s$^{-1}$ for a $kT=0.3{\rm\,keV}$ blackbody and $1.2 \times 10^{-13}$~ergs~cm$^{-2}$~s$^{-1}$ for a power law with $\Gamma =1$. This is then converted to an upper limit on the 0.2--10~keV luminosity of $5.2\times10^{31}{\rm\,ergs\,s^{-1}}$ or $1.3\times10^{32}{\rm\,ergs\,s^{-1}}$, respectively. We discuss the implications for the X-ray counterpart in Section~\ref{sec:multiwavelength_cp}.

Concurrently, {\it Swift} provided Ultra-violet Optical Telescope (UVOT) data with the UVM2 filter centered on 2246\AA. Unfortunately, only 1508\,s of the 10\,ks were available. No detections were made in the VLA localization region. We derive a
5-$\sigma$ upper limit of $m_{AB}>21.69$.

\bibliography{J1634+44}{}
\bibliographystyle{aasjournal}



\end{document}